\documentclass[fleqn,usenatbib]{mnras}
\usepackage{newtxtext,newtxmath}
\usepackage[T1]{fontenc}
\DeclareRobustCommand{\VAN}[3]{#2}
\let\VANthebibliography\thebibliography
\def\thebibliography{\DeclareRobustCommand{\VAN}[3]{##3}\VANthebibliography}

\usepackage{graphicx}	% Including figure files
\usepackage{natbib}
\usepackage{academicons}
\usepackage{hyperref}
\usepackage{xcolor}
\newcommand{\msun}{M_\odot}
\newcommand{\fexsitu}{$f_{\rm ex\,situ}$}
\newcommand{\orcid}[1]{\href{https://orcid.org/#1}{\includegraphics[width=8pt]{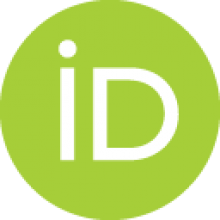}}}

\title[]{On the formation of massive quiescent galaxies with diverse morphologies in the TNG50 simulation}

\author[Park et al.]{Minjung Park$^{1}$\thanks{E-mail: minjung.park@cfa.harvard.edu} \orcid{0000-0002-8435-9402},
Sandro Tacchella$^{2}$
\orcid{0000-0002-8224-4505},
Erica J. Nelson$^{3}$,
Lars Hernquist$^{1}$,
Rainer Weinberger$^{4}$,
\newauthor 
Benedikt Diemer$^{5}$,
Dylan Nelson$^{6}$,
Annalisa Pillepich$^{7}$,
Federico Marinacci$^{8}$
\orcid{0000-0003-3816-7028},
and Mark Vogelsberger$^{9}$
\\
\\
$^{1}$Center for Astrophysics $\mid$ Harvard \& Smithsonian, 60 Garden St., Cambridge, MA 02138, USA\\
$^{2}$Department of Physics, Ulsan National Institute of Science and Technology (UNIST), Ulsan 44919, Republic of Korea\\
$^{3}$Department for Astrophysical and Planetary Science, University of Colorado, Boulder, CO 80309, USA\\
$^{4}$Canadian Institute for Theoretical Astrophysics, 60 St. George St., Toronto, ON M5S 3H8, Canada\\
$^{5}$Department of Astronomy, University of Maryland, College Park, MD 20742, USA\\
$^{6}$Zentrum f{\"u}r Astronomie der Universit{\"a}t Heidelberg, ITA, Albert-Ueberle-Str. 2, D-69120 Heidelberg, Germany\\
$^{7}$Max-Planck-Institut fur Astronomie, K{\"o}nigstuhl 17, 69117 Heidelberg, Germany\\
$^{8}$Department of Physics and Astronomy
``Augusto Righi'', University of Bologna, I-40129 Bologna, Italy \\
$^{9}$Kavli Institute for Astrophysics and Space Research, Massachusetts Institute of Technology, Cambridge, MA 02139, USA
}

\pubyear{2021}

\begin{document}
\label{firstpage}
\pagerange{\pageref{firstpage}--\pageref{lastpage}}
\maketitle

\begin{abstract}
Observations have shown that the star-formation activity and the morphology of galaxies are closely related, but the underlying physical connection is not well understood. Using the TNG50 simulation, we explore the quenching and the morphological evolution of the 102 massive quiescent galaxies in the mass range of $10.5<\log(M_{\rm stellar}/M_{\odot})<11.5$ selected at $z=0$. 
The morphology of galaxies is quantified based on their kinematics, and we measure the quenching timescale of individual galaxies directly from star formation history. 
We show that galaxies tend to be quenched more rapidly if they: (i) are satellites in massive halos, (ii) have lower star-forming gas fractions, or (iii) inject a larger amount of black hole kinetic feedback energy. 
By following the global evolutionary pathways, we conclude that quiescent discs are mainly disc galaxies that are recently and slowly quenched. 
Approximately half of the quiescent ellipticals at $z=0$ are rapidly quenched at higher redshifts while still disc-like.  While being quiescent, they gradually become more elliptical mostly by disc heating, yet these ellipticals still retain some degree of rotation. 
The other half of quiescent ellipticals with the most random motion-dominated kinematics build up large spheroidal components before quenching primarily by mergers, or in some cases, misaligned gas accretion. However, the mergers that contribute to morphological transformation do not immediately quench galaxies in many cases. In summary, we find that quenching and morphological transformation are decoupled. We conclude that the TNG black hole feedback -- in combination with the stochastic merger history of galaxies -- leads to a large diversity of quenching timescales and a rich morphological landscape. 
\end{abstract}


%%%%%%%%%%%%%%%%%%%%%%%%%%%%%%%%%%%%%%%%%%%%%%%%%%

%%%%%%%%%%%%%%%%% BODY OF PAPER %%%%%%%%%%%%%%%%%%
\section{Introduction} \label{sec:intro}
It has long been suggested that the morphology of galaxies is closely related to their star formation activity. 
Many extra-galactic surveys have reported that local galaxies are composed of two distinct populations, star-forming and quiescent galaxies, which produce the galaxy colour bimodality \citep[e.g.,][]{Strateva2001, Baldry2004}.
The distinct colours and star formation rates are often associated with the relative importance of the two most distinctive structures in a galaxy, a disc and a bulge component, as the disc tends to be blue and star-forming, while the bulge component is red and composed of older stars. 
Indeed, many studies have shown that the quiescence of galaxies is connected to larger bulge components or higher central mass densities \citep[e.g.,][]{Wuyts2011, Cheung2012, vanDokkum2014, Mosleh2017, Whitaker2017}. 
Also, the fraction of the quiescent galaxies at $z=0$ has been found to increase with the stellar mass of the galaxies \citep[e.g.,][]{Peng2010MassFunction, Bluck2014, Lang2014BulgeCandels/3D-HST}, suggesting that the bulk population of massive galaxies becomes more bulge-dominated and quiescent with cosmic time.

The large bulge component is thought to be related to the quiescence of galaxies. The bulge components are thought to grow by mergers \citep[e.g.,][]{Toomre1977, Barnes1988ApJ331, Hernquist1992ApJ400, Hernquist1993ApJ409, Hopkins2010MergersMatter} or disc instability \citep[e.g.,][]{Noguchi1999EarlyDisks, Dekel2009FormationSpheroids} that could destroy discs and induce AGN feedback and subsequent quenching \citep{Springel2005BH}. This seems to suggest that the morphological transformation from discs to ellipticals (spheroids) could accompany the quenching process of galaxies \citep{Hopkins2006ApJS163, Hopkins2008ApJS175}.
However, unlike the present-day star-forming galaxies, which are mostly disc-dominated, the morphology of the star-forming galaxies at high redshifts is not very well known. Therefore, it is not yet clear how much bulge grows as galaxies become quiescent and when the bulge growth occurs relative to the quenching: before, during, or after quenching. 
Also, the existence of quiescent (passive or red) disc galaxies, or often referred to as S0s or lenticulars \citep[e.g.,][]{Masters2010}, and the star-forming/blue ellipticals/spheroids \citep[e.g.,][]{Zolotov2015CompactionNuggets, Tacchella2016} has further added complexity to this picture, suggesting that there is no single pathway through which galaxies are quenched and transform their morphology \citep[e.g.,][]{Park2019, Tacchella2019}.

Regarding quenching processes, several mechanisms have been proposed and are thought to work together to quench galaxies. Broadly, the quenching processes can be divided into two classes: (i) preventing gas in a galaxy from forming stars and (ii) removing gas from a galaxy. The first process (i) includes the so-called halo mass quenching, which suggests that accreted gas in a massive halo, with a mass above a critical mass of $\sim 10^{12}\,M_{\odot}$, could be heated by virial shock heating \citep[e.g.,][]{White1978, Birnboim2003, Keres2005}. For long-term quenching, heating the circumgalactic gas and preventing further cooling is needed \citep[e.g.,][]{Nelson2018FirstBimodality, Terrazas2020, Zinger2020}, which can be achieved by stellar and/or (thermal) active galactic nuclei (AGN) feedback \citep[e.g.,][]{Croton2006, Somerville2008, Somerville2015}. Furthermore, the existence of a large bulge component could stabilize the galactic disc against local disc instability, thus, reducing the star formation efficiency. This is referred to as the morphological quenching, as suggested by \cite{Martig2009}.
Finally, the gas removal process (ii) could be driven by the (kinetic) AGN feedback \citep[e.g.,][]{Silk1998} or environmental effects such as ram-pressure stripping \citep[e.g.,][]{Gunn1972}.

There is growing evidence that various quenching processes act in galaxies on two different timescales, ``fast'' and ``slow'' \citep[e.g.,][]{Wu2018, Belli2019, Wild2020}. 
Theoretically, the gas removal process is thought to shut down star formation quite rapidly, whereas star formation is believed to be halted rather slowly by other processes \citep[e.g.,][]{Trussler2020}.
Indeed, many studies have attempted to constrain the quenching timescales of observed galaxies using various methods and showed that quiescent galaxies have a wide range of quenching timescales \citep[e.g.,][]{Tacchella2021}.  
Using the {\sc SIMBA} simulation, \cite{Rodriguez-Montero2019} have shown that their simulated galaxies show a bimodal distribution in quenching timescales (fast and slow).
Some studies have also suggested that galaxies with different morphology are quenched on different timescales; disc galaxies tend to be quenched slowly (slow quenching), while elliptical galaxies, or the slow rotators, are quenched quite rapidly (fast quenching) \citep[e.g.,][]{Schawinski2014, Smethurst2018}. 
This suggests that quiescent disc and elliptical galaxies are formed by different quenching mechanisms acting on different timescales.

In this paper, we study massive galaxies with a mass range of $10.5<\log(M_{\rm stellar}/M_{\odot})<11.5$ in the TNG50 simulation \citep[][]{Pillepich2019, Nelson2019}.
In this mass range, observed morphological properties are well reproduced in TNG \citep[e.g.,][]{Genel2018, Rodriguez-Gomez2019, Tacchella2019}.
A number of studies have explored the quenching of massive galaxies in the IllustrisTNG model (hereafter TNG) adopted in TNG50. For example, \cite{Weinberger2018} have shown that different feedback channels (e.g. stellar vs. AGN) are dominant at different stellar mass regimes at different cosmic epochs: for galaxies with stellar masses above $\log(M_{\rm stellar}/M_{\odot})\sim10.5$, the energy injection via kinetic AGN feedback dominates at lower redshifts. In fact, within the TNG model, a red, quiescent population of high-mass galaxies that is consistent with observations develops only when the SMBH-driven winds get activated \citep{Weinberger2017MNRAS465, Nelson2018FirstBimodality, Donnari2019}. \cite{Weinberger2018} also showed that the star formation efficiency significantly drops for galaxies with a SMBH mass greater than $\sim2\times10^8\,\msun$: the kinetic AGN feedback has been demonstrated to be the key driving factor for quenching in massive galaxies in the TNG model, via both ejecting the gas and preventing it from cooling, via heating \citep{Terrazas2020, Zinger2020, Davies2021}. Also, \cite{Donnari2021a} have studied the quenched fractions of central and satellite galaxies in the TNG100 and TNG300 simulations \citep[][]{Nelson2019ComAC} and concluded that massive galaxies with stellar mass above a few times of $10^{10}\,\msun$, even for satellite galaxies, are quenched by internal processes, most likely by AGN feedback, not by environmental effects.

Whereas it is clear that AGN feedback plays the primary role in quenching massive galaxies in the TNG models \citep[see also][]{Nelson2019, Luo2020WhatGalaxies}, it is still not yet well understood how the morphological evolution is related to star formation history, which will be explored in this work.
Previously, \cite{Joshi2020TheClusters} have shown with TNG50 and TNG100 that the long-lasting reduction of available gas and the mechanisms perturbing preexisting disc stars into non-disc orbits are the two necessary conditions for the long-term morphological transformation from discs to ellipticals. These are achieved by a combination of mergers and AGN feedback for massive field galaxies and by a combination of ram-pressure stripping and tidal shocking at pericentric passages for satellite galaxies.
In this paper, we expand upon these results and aim to answer the following questions by inspecting massive TNG50 galaxies:
\begin{itemize}
	\item Do the quiescent TNG50 galaxies show different (fast and slow) quenching timescales? How does the quenching timescale differ for galaxies with different morphologies (discs vs. ellipticals)? 
	\item If the TNG50 quiescent galaxies also show a variety of quenching timescales, what determines the quenching timescales? 
	\item How does the morphology evolve during and after quenching? Can we understand the formation of quiescent disc and elliptical galaxies?
\end{itemize}

%Add outline of paper.
This paper is structured as follows. In Section~\ref{sec:methodology}, we give a brief description of the TNG50 simulation and describe how we select the quiescent galaxies and how we measure the two most important parameters in our study: the morphology of galaxies and their quenching timescales. In Section~\ref{sec:results}, we present our results on the quenching timescales and the time since quiescence of the disc and elliptical galaxies (Section~\ref{sec:quenching_quiescent_time}) and how quenching timescales depend on various properties such as halo mass, star-forming gas fraction, and AGN feedback energy (Section~\ref{sec:what_affects_quenching_timescales}). In Section~\ref{sec:morphological_evolution}, we explore when the morphologies of galaxies change in the process of quenching (i.e., before, during, or after quenching) to understand the formation of quiescent disc and elliptical galaxies. 
In Section~\ref{sec:discussion}, we discuss how massive quiescent galaxies we observe today could have a variety of morphologies (Section~\ref{sec:discuss_variety_of_morphology}) and present some caveats of our study (Section~\ref{sec:discuss_caveats}). Finally, we summarize our results and conclusions in Section~\ref{sec:summary}.

%%%%%%%%%%%%%%%%%%%%%%%%%%%%%%%%%%%%%
%            Methodology            %
%%%%%%%%%%%%%%%%%%%%%%%%%%%%%%%%%%%%%
\section{Methodology} \label{sec:methodology}

\subsection{The TNG50 simulation} \label{sec:simulation}

The TNG50 simulation\footnote{https://www.tng-project.org} \citep{Pillepich2019,Nelson2019} is a high-resolution magneto-hydrodynamical cosmological volume simulation run using the {\sc Arepo} code, as part of the Illustris TNG project 
\citep{Springel2018MNRAS475, Naiman2018MNRAS477, Marinacci2018MNRAS480, Pillepich2018FirstGalaxies, Nelson2018FirstBimodality, Nelson2019ComAC}, the next generation of the origial Illustris simulation 
(\citealt{Vogelsberger2014IntroducingUniverse}; \citealt{Vogelsberger2014Natur509}; See also \citealt{Vogelsberger2020Nat2}, for review of cosmological simulations).
Here we give a brief description of the simulation.
The box size of the simulation is $\sim 50\,\rm Mpc$, which is the smallest volume among the TNG suite, but with highest resolution. The baryonic and dark matter (DM) mass resolution is $m_{*}\sim 8.5\times10^4\,M_\odot$ and $m_{\rm DM}\sim4.5\times10^5\,M_\odot$, respectively, and and the minimum gravitational softening for the star-forming gas cells is $\sim74\,\rm pc$, with an average cell size for the solution of the magneto-hydrodynamics (MHD) of $100-140\,\rm pc$ throughout the cosmic time.
The high resolution of TNG50 allows us to track the star formation history and the morphology of individual galaxies more reliably for our proposed study.
We used the python framework hydrotools \citep{Diemer2017, Diemer2018ApJSHydrotools} to extract the data from the simulation.
There are $\sim360$ galaxies in the range of $10.5<\log(M_{\rm stellar}/M_\odot)<11.5$ at $z=0.0$. While the simulation mostly covers field environments, there are two Virgo cluster-analogues at $z=0$ with virial masses of $\log(M_{\rm 200\,c}/M_\odot)\sim14$. Furthermore, there are 7 massive group-like objects with $\log(M_{\rm vir}/M_\odot)>13.5$.

In the TNG model, stellar particles representing a coeval population of stars formed stochastically in a gas cell where the density is above the threshold density of $n_{\rm}\simeq0.1\,\rm cm^{-3}$ following the Kennicutt-Schmidt relation \citep{Springel2003MNRAS339}. The Chabrier initial mass function \citep[][]{Chabrier2003GalacticFunction} is assumed for each stellar particle and as the stellar population evolves with time, it returns mass and metals into the surrounding medium by AGB winds, and supernovae Type II and Ia. The TNG model assumes stars with masses of $1-8\,\msun$ go through the AGB phase, and the stars with masses in the range of $8-100\,\msun$ explode as Type II supernovae.
More detailed description regarding the stellar feedback can be found in \cite{Pillepich2018TNGmodels}.

A super-massive black hole (SMBH) with a mass of $\sim10^6\,\msun$ is seeded at the potential minimum of a halo that is more massive than $7.4\times10^{10}\,\msun$ and does not yet contain any other SMBHs. Once the SMBH is seeded, it grows either via SMBH-SMBH mergers or by accretion following the Bondi-Hoyle accretion rate.
The SMBH feedback is modelled in two different ways depending on the ratio of the accretion rate to the Eddington rate. If the ratio exceeds the threshold ratio (i.e., in high-accretion-rate mode), defined as $0.002\,(M_{\rm BH}/ 10^8\,\msun)^2$ with a maximum limit of 0.1, thermal energy is continuously and isotropically injected into the surrounding medium at a rate of $\dot{E}_{\rm thermal}=0.02\dot{M}c^2$; the thermal mode (or quasar mode). On the other hand, if the accretion rate $(\dot{M})$ is low (below the threshold ratio), feedback energy is injected in the form of kinetic energy in a pulsed and directed way; the kinetic mode (or wind mode). The rate of this kinetic feedback energy is $\dot{E}_{\rm kinetic}=\epsilon_{\rm f,kin}\dot{M}c^2$ where $\epsilon_{\rm f,kin}$ depends on the kernel-weighted density of the surrounding medium relative to the density threshold for star formation. A More detailed descriptions of the SMBH feedback in the TNG models can be found in \cite{Weinberger2017MNRAS465, Weinberger2018} and \cite{Pillepich2021}.

\subsection{Sample selection}
To understand the link between quenching and morphological transformation, we select quiescent galaxies at $z=0.0$, including both centrals and satellites, and study their evolution since $z=2$. We use 352 galaxies with stellar mass in the range of $10.5<\log(M_{\rm stellar}/M_\odot)<11.5$: this ensures a sufficient number of quiescent galaxies and also more than $\sim4\times10^5$ stellar particles per galaxy. 
Of 352 galaxies in total, 54 galaxies are cluster satellites at $z=0$ with host halo mass in the range of $\log(M_{\rm 200\,c}/M_\odot)=13.5-14.3$.

\begin{figure*}
    \centering
    \includegraphics[width=\textwidth]{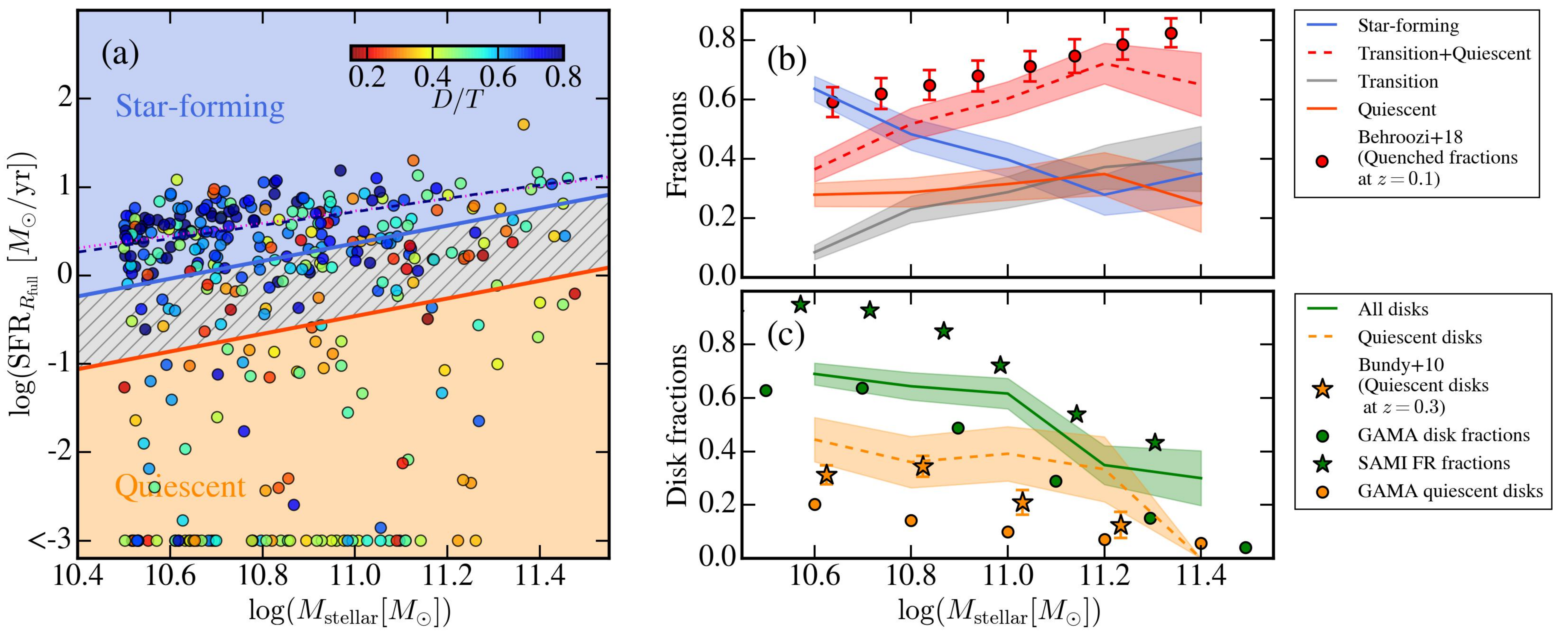}
    \caption{(a) Star-formation rate (SFR) versus stellar mass ($M_{\rm stellar}$) for all TNG50 galaxies in the mass range $10.5<\log(M_{\rm stellar}/M_\odot)<11.5$. Both SFR and stellar mass are measured within $R_{\rm full}$, the radius at which the surface mass density drops below $\Sigma_{*}=10^6\,M_\odot/\rm kpc^2$. Galaxies with sSFR higher than ${\rm sSFR }>1/[3\,t_{\rm H}(z)]$ are classified as star-forming (in the blue region), and galaxies with ${\rm sSFR} <1/[20\,t_{\rm H}(z)]$ are defined as quiescent (orange region). Those in between are considered to be transition galaxies (gray hatched region). The dotted navy and dashed magenta lines are the observed main sequence at $z=0$ from \citet{Whitaker2012} and \citet{Renzini2015}, respectively. The TNG50 star-forming galaxies seem to follow the observed main sequence well. Galaxies are colour-coded by $D/T$, the kinematic morphological indicator. 
    (b) Fractions of star-forming (blue), transition (gray), and quiescent (orange) galaxies as a function of galaxy stellar mass. The fraction of non-star-forming galaxies (sum of transition and quiescent galaxies) is shown by the red dashed line and compared with the observed fractions of passive galaxies from \citet{Behroozi2019} where they adopted data from \citet{Moustakas2013} and \citet{Muzzin2013}. 
    (c) Fractions of disc galaxies ($D/T>0.5$) among all types of galaxies (solid green line) and among the quiescent galaxies (dashed orange line) as a function of stellar mass. The green circles are observed (visually classified) disc fractions in GAMA \citep{Moffett2016GalaxyDiscs}, and the green stars are the fractions of the (kinematically defined) fast-rotators among the SAMI primary sample from \citet{Guo2020}. The orange stars are the visually defined passive disc fractions at $z\sim0.3$ presented in \citet{Bundy2010TheCOSMOS}, and the orange circles are the disc fractions among quiescent galaxies in the GAMA sample (see the text for the selection criterion). 
    We find that the mass trends of the disc fractions (all discs and quiescent discs) in the TNG50 simulation are broadly consistent with observations, although the definitions and methods differ between the two, preventing any strong quantitative comparison.}
    \label{fig:m_sfr}
\end{figure*}

The star-forming and quiescent galaxies can be classified based on a specific star formation rate (sSFR) criteria, as shown in Fig.~\ref{fig:m_sfr} (a). 
Note that both stellar mass and the instantaneous star formation rates (SFRs) are measured within the radius at which the surface stellar mass density drops below $\Sigma_{*}=10^6\,M_{\odot}/\rm kpc^2$ (defined as $R_{\rm full}$), which is an arbitrary value to mimic the surface brightness cut. 
We will discuss the effect of the choice of aperture in due course, but our main conclusions are not affected by the aperture size.
Galaxies are classified as star-forming if ${\rm sSFR }>1/[3\,t_{\rm H}(z)]$ (galaxies in the blue region), where $t_{\rm H}(z)$ is the Hubble time at each redshift, and as quiescent if ${\rm sSFR} <1/[20\,t_{\rm H}(z)]$ (in the orange region). Galaxies between ($ 1/[20\,t_{\rm H}(z)] < {\rm sSFR} <1/[3\,t_{\rm H}(z)]$) are considered transition galaxies (in gray hatched region), making up the green valley.  
There are in total 102 quiescent galaxies in our sample, 30 of them are cluster satellites.

The selected star-forming galaxies seem to follow the main sequence derived from observed galaxies \citep[navy dashed and magenta dotted lines]{Whitaker2012, Renzini2015}. This is consistent with the findings of \cite{Donnari2019}, where they quantified the main sequence of the TNG100 and TNG300 simulations and found minimal resolution effects in the locus of SFMS. 
Also, \cite{Nelson2021} have found that the resolved (on $\sim 1\,\rm kpc$ scales) and integrated SFMS of TNG50 galaxies agree well with observed SFMS from 3D-HST (within 0.1 dex).
The SFR of the quiescent galaxies is about 1.5 dex lower than that of the star-forming galaxies of comparable stellar mass.

Panel (b) of Fig.~\ref{fig:m_sfr} shows the fractions of star-forming (blue line), transition (gray line), and quiescent (orange line) galaxies as a function of stellar mass. The shading of each line indicates the $1\,\sigma$ error of the binomial distribution. The fraction of star-forming galaxies decreases with stellar mass, while the fraction of quiescent galaxies remains almost constant across this mass range. Interestingly, it is the the fraction of transition galaxies that increases with the stellar mass. 
The red dashed line shows the fraction of non-star-forming galaxies (i.e., the sum of transition and quiescent galaxies) which agrees with the observed fractions of ``passive galaxies'' presented in \citet{Behroozi2019} where they adopted data from \cite{Moustakas2013} and \cite{Muzzin2013}. See also \cite{Donnari2019} and \cite{Donnari2021b} for detailed comparison of the TNG quenched fractions to observational results.

\subsection{Kinematic morphology: Disc-to-total ratio (D/T)}

To explore the morphological evolution of individual galaxies, we use the kinematic morphology indicator, disc-to-total ratio, or $D/T$, by selecting ``disc stars'' in a galaxy based on stellar orbits.  The ``circularity'' of stellar orbits, i.e. how close a stellar orbit is to a circular orbit in the galactic midplane, can be calculated as the circularity parameter ($\epsilon$) introduced in \cite{Abadi2003SimulationsDisks}. It is defined as follows: $\epsilon=J_z/J_{\rm circ}$, where $J_z$ is the angular momentum of each stellar particle in the direction of the net rotation axis of a galaxy defined using only the stellar particles within $3\,R_{\rm eff}$ ($R_{\rm eff}$: effective radii, the radius containing half the total stellar mass), and $J_{\rm circ}$ is the angular momentum of a circular orbit in the galactic midplane with the same binding energy as the stellar particle in question.
We take disc stars to be the stellar particles with a circularity parameter greater than 0.5. By measuring how many stars in a galaxy have nearly circular orbits with $\epsilon>0.5$, we can measure $D/T$. Each galaxy in Fig.~\ref{fig:m_sfr} (a) is colour-coded by $D/T$. We then separate galaxies with different morphologies based on $D/T$: the disc ($D/T>0.5$) and elliptical ($D/T<0.5$) galaxies.

Panel (c) of Fig.~\ref{fig:m_sfr} shows the fractions of disc galaxies among all types of galaxies (solid green line) and among the quiescent galaxies (dashed orange line) as a function of stellar mass. The observed fractions of discy galaxies are shown as circles and stars; the green circles are observed (visually classified) disc fractions in GAMA \citep{Moffett2016GalaxyDiscs}, and the green stars are the fractions of the (kinematically defined) fast-rotators among the SAMI primary sample from \cite{Guo2020}. The orange stars are the visually defined passive disc fractions at $z\sim0.3$ presented in \cite{Bundy2010TheCOSMOS}, and the orange circles are the disc fractions among the quiescent galaxies in the GAMA sample where we select disc galaxies to be those with the S\'{e}rsic index measured in the K band lower than 2 and quiescent galaxies using the same sSFR criterion (i.e., ${\rm sSFR} <1/[20\,t_{\rm H}(z)]$).

At face value, this comparison would suggest that TNG50 has a higher fraction of disc galaxies (both all types of discs and quiescent discs) than observed disc fractions which are visually selected. This apparent discrepancy, however, may primarily result from the different definitions of ``discs'', as we used a kinematic morphological indicator to define disc galaxies, while in observations, they are classified based on visual morphology. 
Similarly, the kinematic-based observational measurements also differ in detail from our methods (i.e. projected line-of-sight velocities in data versus full 3D information in the simulation). As a result, we cannot say whether the observed fast-rotator fractions, shown as green stars, are truly higher than the TNG50 fractions, as implied.
%However, when kinematic measurements are included, the observed fast-rotator fractions, shown as green stars, seem to be even higher than our kinematically-defined disc fractions. 
Therefore, while it is difficult to make an one-to-one comparison of the disc fractions in the TNG50 simulation to observations due to the different definitions of ``disc'' galaxies, the qualitative mass trend, where the disc fraction mildly decreases with stellar mass of the galaxies, seems to be consistent with observed trends and those found in TNG100 \citep[e.g.,][]{Tacchella2019}.

\subsection{Quenching timescales and time since quiescence}

The quenching timescale of a galaxy (i.e., how fast galaxies are quenched) can be measured using the same sSFR criteria mentioned above. 
We tracked the main progenitors of a galaxy up to $z\sim2$, and the sSFR of the galaxy at each time is measured within the radius at which the surface mass density drops below $\Sigma_{*}=10^6\,M_{\odot}/\rm kpc^2$ for the galaxy at each period. 
To avoid small fluctuations in sSFR, at each point, the sSFR is taken as the median between 5 nearby snapshots ($\pm$ 2 snapshots in the neighborhood). The time interval for each snapshot is $\sim160\,\rm Myr$. 
We confirm that very rapid quenching (e.g., shorter than the time interval between two consecutive snapshots) could be measured even using this median filtering process.

We also measure the time since quiescence, the amount of time the galaxy has been quiescent, i.e., the time interval between when it becomes quenched and $z=0$.
This tells us whether galaxies are recently quenched (short time since quiescence) or quenched earlier at higher redshifts (longer time since quiescence).
Here we list the definition of the timescales/epochs that we use in this study. 

\begin{itemize}
	\item The epoch of the onset of quenching ($t_{\rm onset}$): it is defined as the epoch at which the sSFR drops to ${\rm sSFR }=1/[3\,t_{\rm H}(z)]$  
	\item The epoch when the quenching ends ($t_{\rm end}$): it is defined as the epoch at which the sSFR drops to ${\rm sSFR }=1/[20\,t_{\rm H}(z)]$  
	\item Quenching timescale: the duration of quenching is defined as the time period during which the sSFR drops from ${\rm sSFR }=1/[3\,t_{\rm H}(z)]$ to $1/[20\,t_{\rm H}(z)]$ (i.e., $t_{\rm end} - t_{\rm onset}$)
	\item Time since quiescence: it is defined as the amount of time the galaxy has been quiescent until $z=0$ (i.e., $t_{z=0} - t_{\rm end}$)
\end{itemize}

%The quenching timescale of a galaxy, i.e., how fast galaxies are quenched, can be measured using the same sSFR criteria mentioned above. 
%We define the onset epoch of quenching ($t_{\rm onset}$) as the epoch at which the sSFR drops to ${\rm sSFR }=1/[3\,t_{\rm H}(z)]$.  
%The epoch when the quenching ends ($t_{\rm end}$) is defined when the sSFR drops to ${\rm sSFR }=1/[20\,t_{\rm H}(z)]$. 
%The quenching timescale (the duration of quenching) is then defined the time period during which the sSFR drops from ${\rm sSFR }=1/[3\,t_{\rm H}(z)]$ to $1/[20\,t_{\rm H}(z)]$.
%The sSFR at each time is measured within the radius at which the surface mass density drops below $\Sigma_{*}=10^6\,M_{\odot}/\rm kpc^2$ for the galaxy at each period. 
%To avoid small fluctuations in sSFR, at each point, the sSFR is taken as the median between 5 nearby snapshots ($\pm$ 2 snapshots in the neighborhood). The time interval for each snapshot is $\sim160\,\rm Myr$. 
%We confirm that very rapid quenching (e.g., shorter than the time interval between two consecutive snapshots) could be measured even using this median filtering process.
%We also measure the time since quiescence, the amount of time the galaxy has been quiescent, i.e., the time interval between when it becomes quenched and $z=0$.
%This tells us whether galaxies are recently quenched (short time since quiescence) or quenched earlier at higher redshifts (longer time since quiescence).

%%%%%%%%%%%%%%%%%%%%%%%%%%%%%%%%%%%%%
%              Results              %
%%%%%%%%%%%%%%%%%%%%%%%%%%%%%%%%%%%%%
\section{Results} \label{sec:results}

\subsection{Quenching timescales and time since quiescence for disc and elliptical galaxies} \label{sec:quenching_quiescent_time}

\begin{figure*}
    \centering
    \includegraphics[width=0.7\textwidth]{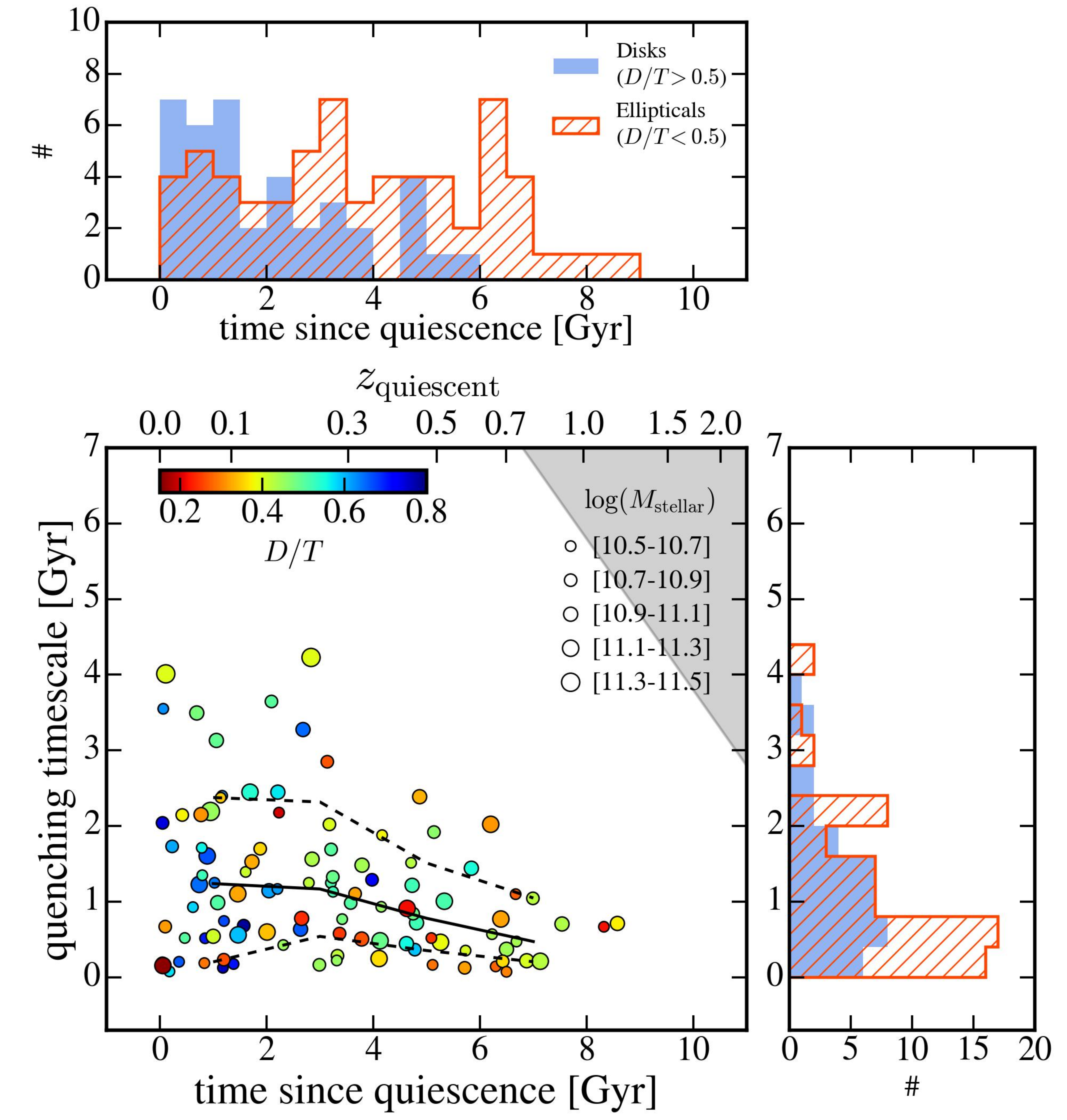}
    \caption{The quenching timescales and time since quiescence for quiescent TNG50 galaxies selected at $z=0$. Each galaxy is colour-coded by $D/T$ at $z=0$, and the sizes of the circles indicates the stellar mass of the galaxies (larger circles for more massive galaxies). The black solid line shows the median quenching timescale at each epoch and the lower/upper dashed line indicates 16th/84th percentiles, which suggests a mild trend of increasing quenching timescale with decreasing redshift. The gray shade on the upper right corner of the main panel indicates the regions where the time spent since the onset of quenching (i.e., sum of quenching timescales and time since quiescence) is greater than the age of the universe. The upper/right panel shows the distributions of quenching timescales/time since quiescence for quiescent disc (blue solid histogram) and elliptical (orange hatched) galaxies classified based on $D/T=0.5$. We find that quiescent disc and elliptical galaxies show different distributions in quenching timescales and time since quiescence; quiescent discs are mostly recently and slowly quenched. On the other hand, quiescent ellipticals are mostly quenched more rapidly and they have a wide range of time since quiescence (both recently quenched and quenched long ago).}
    \label{fig:quenching_quiescent_time_DT}
\end{figure*}

Fig.~\ref{fig:quenching_quiescent_time_DT} shows the quenching timescale (i.e., how rapidly galaxies are quenched) and the time since quiescence (i.e., how long galaxies have been quiescent) for quiescent galaxies at $z=0$. Each galaxy is colour-coded by $D/T$ at $z=0$, and the size of markers indicates the stellar mass of the galaxies (larger circles for more massive galaxies). Based on the distribution of galaxies with different masses (circles of different sizes), it seems that there is no clear mass dependence for the quenching timescales and time since quiescence.

The present-day massive quiescent galaxies in TNG50 have become quiescent since $z\lesssim1$\footnote{Note that we track the star formation history of quiescent galaxies with stellar mass of $\log(M_{\rm stellar}/M_\odot) \sim10.5-11.5$ at $z=0.0$. Therefore, we did not consider the quiescent galaxies that emerged before $z\sim1$ but evolved to become more massive than $\log(M_{\rm stellar}/M_\odot)>11.5)$ by $z=0$.} (with time since quiescence of $\lesssim 9\,\rm Gyr$). 
Also, TNG50 quiescent galaxies show a wide range of quenching timescales between $0-4\,\rm Gyr$. 
The black solid line shows the median quenching timescales at each epoch, and the lower/upper dashed line indicates 16th/84th percentiles. There seems to be a mild trend of increasing quenching timescale with decreasing redshifts; in other words, quiescent galaxies at higher redshifts (e.g., $z>0.7$) are quenched rapidly (e.g., $<1\,\rm Gyr$), whereas, there are both rapidly and slowly-quenched galaxies at lower redshifts. This redshift-dependent trend of quenching timescale agrees with the results found in other studies based on the change in number density of transition populations \citep[e.g.,][]{Pandya2017}.

The upper panel shows the distribution of time since quiescence for disc (blue solid histogram) and elliptical (orange hatched histogram) galaxies divided based on $D/T$ at $z=0$. While quiescent disc galaxies are recently quenched (mostly within the last $2\,\rm Gyr$), quiescent elliptical galaxies at $z=0$ have a much wider range of time since quiescence; galaxies that have been quiescent for a long time (e.g., time since quiescence $>6\,\rm Gyr$ since $z=0.7$) are all ellipticals at $z=0$. This morphological dependence on time since quiescence is consistent with the results found in \cite{Correa2019}, in which they measured when disc and elliptical galaxies in the EAGLE simulation joined the red sequence based on colour evolution.

Also, disc and elliptical galaxies at $z=0$ show different distributions of quenching timescales, as shown in the right panel. While most of the elliptical galaxies are quenched quite rapidly (57\% of them have quenching timescales shorter than $1\,\rm Gyr$ and 32\% of them have timescales even shorter than $0.5\,\rm Gyr$.), disc galaxies seem to be quenched more slowly.
Several studies have also shown a similar trend that elliptical galaxies are quenched on short timescales while disc galaxies tend to be quenched slowly \citep[e.g., ][]{Schawinski2014, Smethurst2018}.

\cite{Nelson2018FirstBimodality} modelled optical colours of galaxies in TNG100 and TNG300, taking into account attenuation from interstellar dust, and measured the transition timescale ($\Delta t_{\rm green}$) from blue to red sequence, which is a similar quantity as the quenching timescales in this work. The median transition timescale of quiescent galaxies at $z=0$ is $\Delta t_{\rm green}\sim1.6\,\rm Gyr$, which is slightly longer than the median quenching timescale of our sample ($\sim 0.95\,\rm Gyr$).
Interestingly, while we show that disc and elliptical galaxies have different distribution of quenching timescales, they found no correlation between the final morphology and $\Delta t_{\rm green}$. 
On the other hand, they found that there is a decreasing trend of $\Delta t_{\rm green}$ with the stellar mass at the arrival into the red sequence, while we do not see a clear trend with stellar mass, as indicated by the size of markers in Fig.~\ref{fig:quenching_quiescent_time_DT}. 
This may be due to the fact that the mass range of our samples is much narrower than that of their samples in \cite{Nelson2018FirstBimodality}, and the difference in the selection of quiescent/red galaxies, even with different central/satellite ratios depending on the simulated volume. It could also be due to the fact that sSFR is not necessarily linear with colours, as colours are more sensitive to the presence of hot/massive stars.

\subsection{What affects the quenching timescales?}
\label{sec:what_affects_quenching_timescales}

\begin{figure*}
    \centering
    \includegraphics[width=0.95\textwidth]{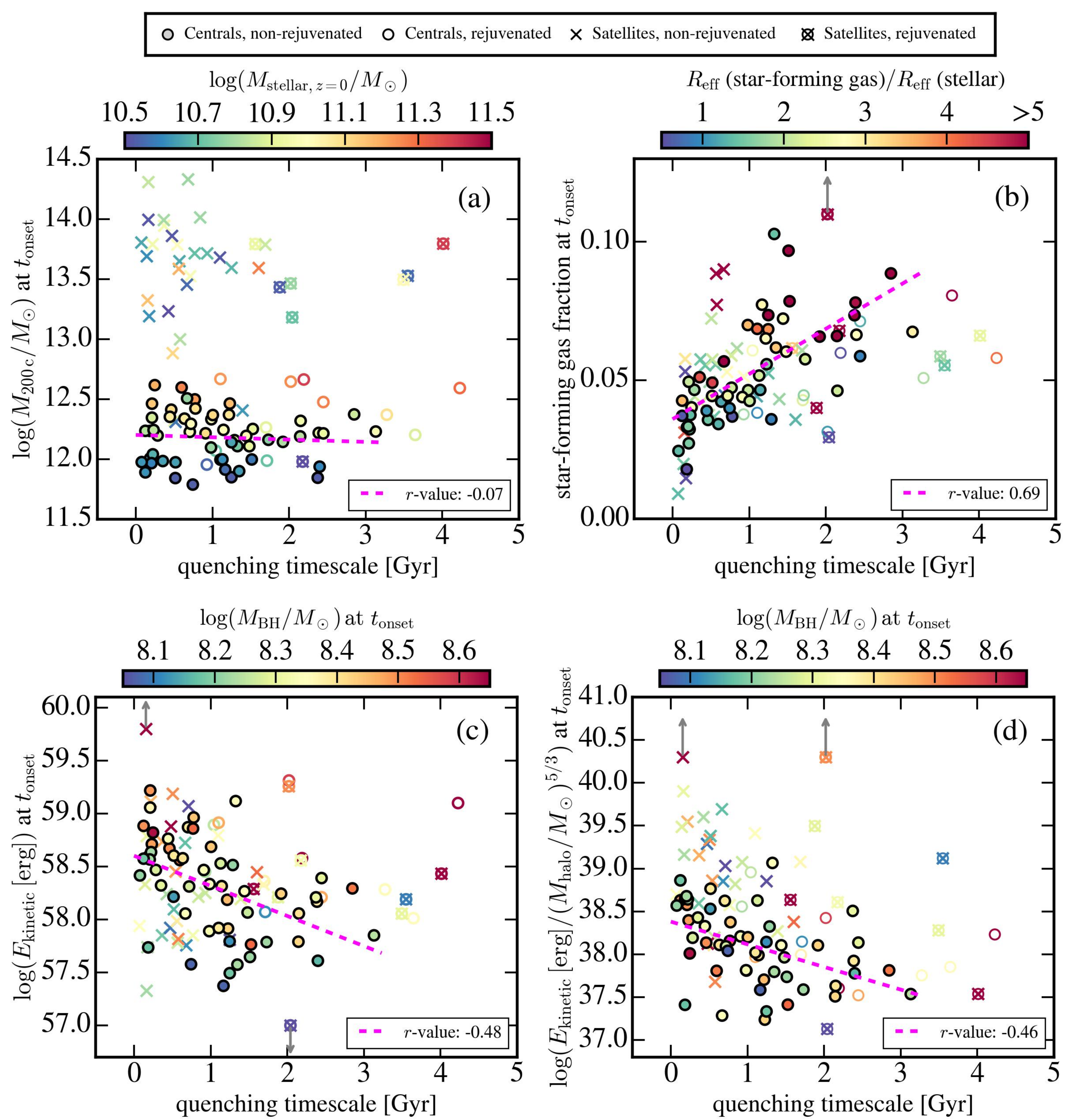}
    \caption{The properties at the onset of quenching as a function of quenching timescales for quiescent TNG50 galaxies selected at $z=0$: (a) the host halo (group) virial mass ($M_{200\,\rm c}$), (b) star-forming gas fraction, (c) AGN feedback energy in the kinetic mode, (d) AGN feedback energy in the kinetic mode normalized by $M_{\rm halo}^{5/3}$ (where $M_{\rm halo}$ is the individual subhalo mass) as a proxy for the gravitational binding energy of each galaxy. 
    Galaxies that are satellites when the quenching begins are marked as crosses, and the galaxies that are rejuvenated (see the text for the definition) during the quenching are shown as open circles. Only the non-rejuvenated central galaxies (filled circles) are used to find the trends between the properties at the onset of quenching and the quenching timescales, which are shown as magenta lines and the corresponding correlation coefficient in the lower right corners.
    The colour code indicates: (a) the stellar mass of galaxies at $z=0$, (b) the ratio of the effective radii between the star-forming gas and the stellar components at the onset of quenching, and the SMBH mass at the onset of quenching for both panels (c) and (d). 
    We find that galaxies tend to be quenched more rapidly (shorter quenching timescales) if they are satellites in a massive halo, have lower star-forming gas fractions, and/or release higher AGN feedback energy at the onset of quenching.}
    \label{fig:onset_props_quenching_time}
\end{figure*}

As described in the introduction, various quenching mechanisms are acting on galaxies on different timescales.
In this section, we investigate the internal properties of galaxies at the onset of quenching to see what affects the quenching timescales of galaxies. 
To provide a fair comparison of the link between the internal properties and the quenching timescales, we identify two external effects that could affect the quenching timescales.
One is an environmental effect, such as harassment or ram-pressure stripping \citep[e.g.,][]{Gunn1972}, which may play a role in effectively removing gas from satellite galaxies in large clusters \citep[e.g.,][]{Jung2018OnSimulations, Yun2019MNRAS483}. 
Indeed, many studies have reported that satellite galaxies show higher quenched fractions than centrals \citep[e.g.,][]{Bluck2014}, implying that they have experienced additional external quenching processes.
Therefore, in Fig.~\ref{fig:onset_props_quenching_time}, they are denoted by crosses (``$\times$'') and separated from those that are quenched as centrals (filled circles).

Another is additional external cold gas accretion during the quenching process, which can feed star formation \citep[e.g.,][]{Keres2005, Dekel2009ColdFormation}, allowing galaxies to have extended quenching timescales.
Therefore, we identify these galaxies with longer quenching timescales due to the non-vanishing presence of star-forming interstellar medium (ISM) gas, as follows.
%with longer quenching timescales due to cold gas accretion.
We first measure the time duration (within the quenching duration) when the sSFR increases or remains the same, and if this period lasts for at least $0.5\,\rm Gyr$, we identify these galaxies as ``weakly rejuvenated'' galaxies during quenching. 
Note that the term ``weak rejuvenation'' used in this study might differ from that used in the literature, as an increase in sSFR during quenching cannot be directly observed.

There are in total 7 weakly-rejuvenated galaxies among the 39 quiescent discs, and 12 out of 63 quiescent ellipticals are identified as weakly rejuvenated.
However, only two galaxies in our sample (both are elliptical at $z=0$) are rejuvenated strongly enough to be re-classified as star-forming galaxies and then quenched again to become quiescent. 
In this sense, galaxies that have been strongly rejuvenated are quite rare in our quiescent galaxy sample. 
%strongly rejuvenated galaxies are quite rare in our sample.
In Fig.~\ref{fig:onset_props_quenching_time}, we mark the ``weakly rejuvenated'' galaxies as open circles and exclude them from analysing the overall trend. 
Note that some satellite galaxies (marked as crosses) could be also rejuvenated during quenching, in which case, they are marked as open circles as well as crosses (thus, '$\bigotimes$').
However, there are not many satellite galaxies that are rejuvenated during quenching. 
The key for the classification is given at the top of Fig.~\ref{fig:onset_props_quenching_time}.

\subsubsection{Central versus Satellites}

Fig.~\ref{fig:onset_props_quenching_time}(a) shows the virial mass of the groups (host halo mass, $M_{\rm 200\,c}$) to which galaxies belong when they begin to be quenched (i.e., at the onset of quenching, denoted by $t_{\rm onset}$) as a function of the quenching timescales. Galaxies quenched as satellites inside large groups with a virial mass of $\log(M_{\rm 200\,c}/M_\odot) \sim13-14$ (denoted as ``$\times$''), mostly have short quenching timescales probably because of environmental effects that could remove the gas.
The short quenching timescales for satellite galaxies in TNG50 \citep[e.g.,][]{Joshi2020TheClusters} ($<1\,\rm Gyr$) agree with the quenching timescales inferred from observational datasets and models \citep[e.g.,][]{Wetzel2013, Rhee2020}. 
%predicted in many other previous work 

Central and non-rejuvenated galaxies are marked as filled circles colour-coded by the stellar mass at $z=0$. 
Their quenching timescales do not seem to have a clear dependence on stellar mass, but they start to be quenched when their host halo mass is around $M_{\rm 200\,c}\sim10^{12}\,M_\odot$, which corresponds to the halo mass where efficient SMBH feedback in the TNG model begins to quench the galaxy population \citep{Weinberger2018, Nelson2018FirstBimodality}.

%nelson+18: relation 

\subsubsection{Star forming gas fractions}
The cold and dense molecular gas is a major ingredient for star formation, thus, quenching has been conventionally associated with the lack of this fuel. 
However, recent observational studies have reported that some of the post-starburst or quiescent galaxies have large molecular gas reservoirs \citep[e.g.,][]{French2015, Suess2017, Hunt2018StellarZ0.7}, suggesting that galaxies could be quenched when the cold molecular gas in the galaxies is inefficient at forming stars. 
Several mechanisms seem to be responsible for the low SF efficiency, including the presence of bulges, the so-called morphological quenching, \citep[e.g.,][]{Martig2009}, or bars \citep[e.g.,][]{Khoperskov2018}.

In the TNG simulations, molecular gas is not self-consistently modeled, thus, we instead measure the mass of the gas in the cells where the star formation is occurring to investigate the relationship between the amount of star-forming gas and the quenching timescales. 
The star-forming gas mass is also measured within the same aperture, $R_{\rm full}$, as are the SFR and $M_{\rm stellar}$.
In Fig.~\ref{fig:onset_props_quenching_time} (b), we measure the fraction of star-forming gas in the galaxies ($=M_{\rm SF\, gas}$/$M_{\rm stellar}$) at the onset of quenching. 
Also, galaxies are colour-coded by the ratio between the effective radius of the star-forming gas and that of the stellar component to see how the distribution of the star-forming gas with respect to the stellar distribution affects the quenching timescales. The effective radius of each component is measured as the radius within which it contains half of the total mass defined as the mass within $R_{\rm full}$.

We find that galaxies having a higher fraction of star-forming gas when they start to be quenched have longer quenching timescales. The magenta dashed line shows a linear fit to the central, non-rejuvenated galaxies (filled circles), and the resulting correlation coefficient of the fit is $\sim$0.69. 
Also, galaxies with higher star-forming gas fractions tend to have more extended distributions of star-forming gas compared to the stellar distributions (redder colours).
%It takes much longer for galaxies to be quenched when the star-forming gas in the galaxies is more widely distributed. 
The high ratios between the effective radii can be mainly explained by two cases; (i) many of the galaxies having higher ratios are accreting gas from the outskirts after mergers, (ii) while star-forming gas is mostly distributed in the disc regions, the stellar component is very compact.
On the other hand, most galaxies with short quenching timescales, shorter than $\rm 1\,Gyr$, have low star-forming gas fractions when the quenching begins, and in those galaxies, star-forming gas is more concentrated in the central regions (lower $R_{\rm eff}$ ratios between the star-forming gas and the stellar components).

\subsubsection{Kinetic AGN feedback}

AGN feedback has been proposed as a mechanism that could quench massive galaxies and form massive quiescent elliptical galaxies \citep[e.g.,][]{DiMatteo2005, Springel2005BH, Croton2006, Dubois2016TheFeedback}.
It is often described as acting in two distinct manners; active SMBHs are believed to inject thermal energy into the surrounding medium and prevent gas from forming stars. Also, they could blow away the gas in the form of outflows, such that galaxies cannot form any more stars as there is no available gas for star formation.
Motivated by this, the AGN feedback prescription in TNG is modeled in two modes, the kinetic (wind-like) and the thermal (quasar) modes, depending on the ratio of the accretion rate to the Eddington rate (see Section~\ref{sec:simulation} and \citealt{Weinberger2017MNRAS465} for more details). 
\cite{Weinberger2018} showed that quenching of the massive (central) TNG galaxies is driven primarily by the kinetic (wind-like mode) AGN feedback, whereas the energy injected via thermal AGN mode is mostly radiated away, at least partially as a consequence of limited numerical resolution, thus, inefficient at quenching galaxies.

Here we measure how much energy is released in the kinetic AGN feedback mode when the quenching begins (averaged over $t_{\rm onset} \pm 100\,\rm Myr$) and explore how it affects the quenching timescales.
In Panel (c) of Fig.~\ref{fig:onset_props_quenching_time}, the kinetic AGN feedback energy released at the onset of quenching is plotted as a function of the quenching timescales, and in panel (d), the AGN feedback energy is normalized by the subhalo mass of each galaxy, $M_{\rm halo}^{5/3}$, as a proxy of the gravitational binding energy. The colour codes in both panels (c) and (d) indicate the SMBH mass of the galaxies at the onset of quenching.

Galaxies in which higher AGN feedback energy is released at the onset of quenching ($\pm 100\,\rm Myr$) seem to be quenched more rapidly; both injected energy in kinetic-mode (in Panel c) and the injected energy normalized by $M_{\rm halo}^{5/3}$ (in Panel d) have a mild decreasing trend with quenching timescales with correlation coefficients of $\sim-0.5$. 
Also, the SMBH mass (indicated as colour code) does not seem to be very tightly correlated with the quenching timescales or the amount of injected feedback energy, as long as the SMBHs are massive enough, e.g., above a few times of $10^8\,\msun$. Several studies have also shown that in the TNG model, quenching is most efficient in galaxies with SMBHs of masses greater than a few times of $10^8\,\msun$ \citep[e.g.,][]{Weinberger2018, Terrazas2020, Quai2021}, and indeed, many of our galaxies have SMBHs with masses above this mass when quenching begins.

\subsection{Morphological evolution in the process of quenching}
\label{sec:morphological_evolution}

In the previous section, we investigated the dependence of quenching timescales on different internal properties, such as star-forming gas fractions and AGN activity. 
We additionally considered whether external effects such as if galaxies are centrals or satellites and if galaxies are rejuvenated, could affect the quenching timescales. 
However, it still remains unclear when and how the morphology of galaxies evolves during quenching and how quiescent galaxies end up having a variety of morphologies.
In this section, we track the evolution of $D/T$ to determine when morphological transformation occurs: before, during, or after quenching.

\subsubsection{The change in $D/T$ during the quenching and quiescent periods}

\begin{figure*}
    \centering
    \includegraphics[width=\textwidth]{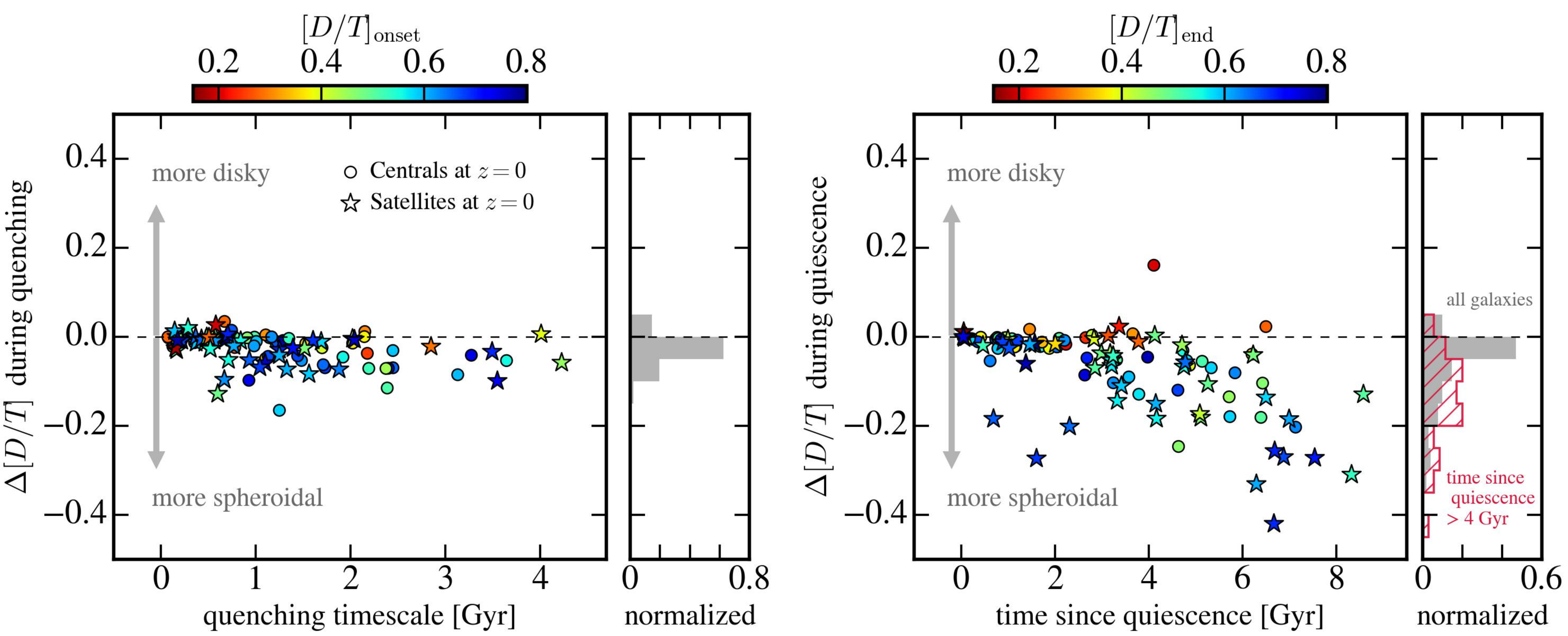}
    \caption{The change in $D/T$ of TNG50 galaxies during quenching (left) and during the time since quiescence (right). Galaxies are colour-coded by the $D/T$ at the onset of quenching (left) and at the time when the quenching is completed (right). Central and satellite galaxies at $z=0$ are shown as circles and stars. The gray histogram right next to each panel shows the $\Delta [D/T]$ distribution normalized by the number of quiescent galaxies, and the red hatched histogram shows only the $\Delta [D/T]$ of the galaxies with time since quiescence longer than $\rm 4\,\rm Gyr$. We find that galaxies do not seem to change their $D/T$ during quenching, and only the galaxies that spent a long time being quiescent (e.g., $>4\,\rm Gyr$) transformed their morphology from discy to elliptical. Also, many of these galaxies that changed their morphology are satellites at $z=0$. 
    }
    \label{fig:DT_evol_during_after_quenching}
\end{figure*}

Fig.~\ref{fig:DT_evol_during_after_quenching} shows the change in $D/T$ for all the present-day quiescent galaxies during their quenching period (between $t_{\rm onset}$, at the onset of quenching, and $t_{\rm end}$, when quenching ends)  
and during the quiescent period between $t_{\rm end}$ and $t_{z=0}$. 
Each galaxy is colour-coded by the $D/T$ from which the change in $D/T$ is measured: $D/T$ at the onset of quenching ($[D/T]_{\rm onset}$, left panel) and at the time when quenching is completed ($[D/T]_{\rm end}$, right panel). Positive $\Delta [D/T]$ means that $D/T$ has increased (galaxies becoming more discy) from the $D/T$ indicated by the colour, while negative $\Delta [D/T]$ means that galaxies become more spheroidal during the period. The circle indicates galaxies that are centrals at $z=0$, while satellites are shown as stars.

We find that galaxies do not seem to change their $D/T$ much during quenching, even for galaxies that have long quenching timescales. The gray histogram on the right in the left panel shows the distribution of $\Delta [D/T]$ during quenching normalized by the total number of quiescent galaxies at $z=0$, and most galaxies are distributed around $\Delta [D/T]\sim0$.
However, the right panel shows that galaxies that spent a long time being quiescent change their $D/T$ significantly; the hatched red histogram on the right shows the $\Delta [D/T]$ for galaxies with time since quiescence greater than $\rm 4\,Gyr$, and they have a wide distribution of $\Delta [D/T]$ centered around $-0.1$.
Also, many of the galaxies that change their morphology dramatically are satellite galaxies marked as stars, possibly because quenched satellite galaxies are more susceptible to gravitational perturbations that could drive the morphological transformation.
%, as they tend to be lower-mass galaxies in the environments where further cold gas accretion is unlikely. 
In conclusion, in contrast to the standard picture, we find that TNG50 galaxies do not change their morphologies significantly during quenching.

\subsubsection{Morphological evolutionary path towards quiescent disc and elliptical galaxies}
\label{sec:evolutionary_parths}
As the colour of the galaxies in Fig.~\ref{fig:DT_evol_during_after_quenching} shows, many of the galaxies started to be quenched with discy morphology (bluer colours) and those disc galaxies that spent a long time being quiescent (e.g., $\rm >4\,Gyr$) change their morphology to become more spheroidal (having lower $D/T$). However, galaxies that were already elliptical before quenching (redder colours in Fig.~\ref{fig:DT_evol_during_after_quenching}) tend to retain their morphology until $z=0$ (only a very few of them become slightly more discy at $z=0$). In this section, we investigate the morphologies of galaxies at the onset of quenching in more detail; how it depends on redshift or how many galaxies started to be quenched with discy/elliptical morphologies, and what determines the morphology at the onset of quenching. Thus, eventually, we aim to identify different formation pathways towards the quiescent disc and elliptical galaxies at $z=0$.

Fig.~\ref{fig:DTevolution_for_discs_ellips} panel (a) shows the morphology of galaxies at the onset of quenching ($[D/T]_{\rm onset}$) versus the time at the onset of quenching ($t_{\rm onset}$).
Galaxies that are centrals at the onset of quenching are plotted as circles, while the stars show satellites at the onset of quenching.
At higher redshifts ($z>0.7$), most galaxies that begin to be quenched are disc galaxies ($[D/T]_{\rm onset}>0.5$). This is probably because most of the spheroidal galaxies at these redshifts are formed as a result of frequent gas-rich mergers and they would be still star-forming \citep[e.g., ][]{Osborne2020} so that there would not be many elliptical galaxies about to be quenched. At lower redshifts, on the other hand, galaxies have diverse morphologies at the onset of quenching; the majority remain discy but there is a distinct minority with  elliptical morphologies. What determines the morphology at the onset of quenching will be discussed in later paragraphs.

We divide quiescent galaxies at $z=0$ into three groups depending on their present-day morphology and their morphology at the onset of quenching and colour the edges of the markers differently: quiescent disc galaxies in blue, quiescent elliptical galaxies with discy morphology at the onset of quenching in orange ($[D/T]_{z=0}<0.5$ and $[D/T]_{\rm onset}>0.5$), and quiescent elliptical galaxies with elliptical morphology when they start to be quenched in dark red ($[D/T]_{z=0}<0.5$ and $[D/T]_{\rm onset}<0.5$). The colours inside the markers indicate the quenching timescales of the galaxies. 
The arrows show the average morphological evolutionary trajectory of the galaxies in the three groups during quenching (hatched arrows) and quiescent periods (solid arrows). Also, the panels on the right show the histograms of the properties of galaxies in the three groups (in blue solid histogram and orange and dark red hatched histograms): (b) present stellar mass, (c) present $D/T$, (d) quenching timescales, and (e) time since quiescence.

For quiescent discs, all of them were discs when they started to be quenched, implying that the rejuvenated scenario for the formation of quiescent discs where elliptical galaxies accrete gas after mergers and form the disc structures \citep[e.g.,][]{Diaz2018, Hao2019} is extremely rare in the TNG50 simulation.

The colours inside the markers show that the quiescent disc galaxies have different quenching timescales depending on when they started to be quenched; galaxies that started to be quenched recently tend to have shorter quenching timescales (otherwise, they would not have been selected as quiescent galaxies at $z=0$, but rather, they would be transition galaxies at $z=0$) and those that started to be quenched at higher redshifts tend to have slightly longer timescales. Thus, the overall time since quiescence (the time that galaxies spent being quiescent until $z=0$) are skewed to shorter timescales, meaning they are mostly quenched recently (Panel e). Also, the majority of quiescent disc galaxies are centrals at the onset of quenching, as plotted as circles in the left panel. However, they tend to be less massive galaxies compared to elliptical galaxies (Panel b).

For quiescent elliptical galaxies at $z=0$, approximately half of them (31/63) have discy morphologies at the onset of quenching and transform their morphology after quenching (plotted as orange). 
They started to be quenched at higher redshifts, and most of them are quenched rapidly ($\rm <1\,Gyr$, panel b). 
Thus, they spend a long time until $z=0$ being quiescent (longer time since quiescence as shown in Panel e) and transform their morphology from discs to ellipticals (ending up having lower $D/T$ at $z=0$).
Many of them are satellite galaxies at the onset of quenching, as marked as stars in the left panel, and at $z=0$, 21 of 31 galaxies are identified as satellites.
They are less massive than the elliptical galaxies starting to be quenched with elliptical morphology (dark red).

The other half (32/63) of the present-day quiescent elliptical galaxies (shown as dark red) already have elliptical morphologies at the onset of quenching and they retain their elliptical morphology until $z=0$ (dark red). They mostly started to be quenched at intermediate/lower redshifts and some of them have notably longer quenching timescales (darker green colours inside markers). Thus, they tend to be the elliptical galaxies recently quenched (shorter time since quiescence in Panel d), i.e., younger elliptical populations at $z=0$. Interestingly, they are overall slightly more massive than the elliptical galaxies that transformed their morphology from discs to elliptical while being quiescent (Panel b). This suggests that their elliptical morphology at the onset of quenching might be related to the mechanism related to their stellar mass growth. We will explore in more detail what determines the morphology of galaxies when they begin to be quenched.

\subsubsection{What determines the morphology of galaxies at the onset of quenching?}

\begin{figure*}
    \centering
    \includegraphics[width=\textwidth]{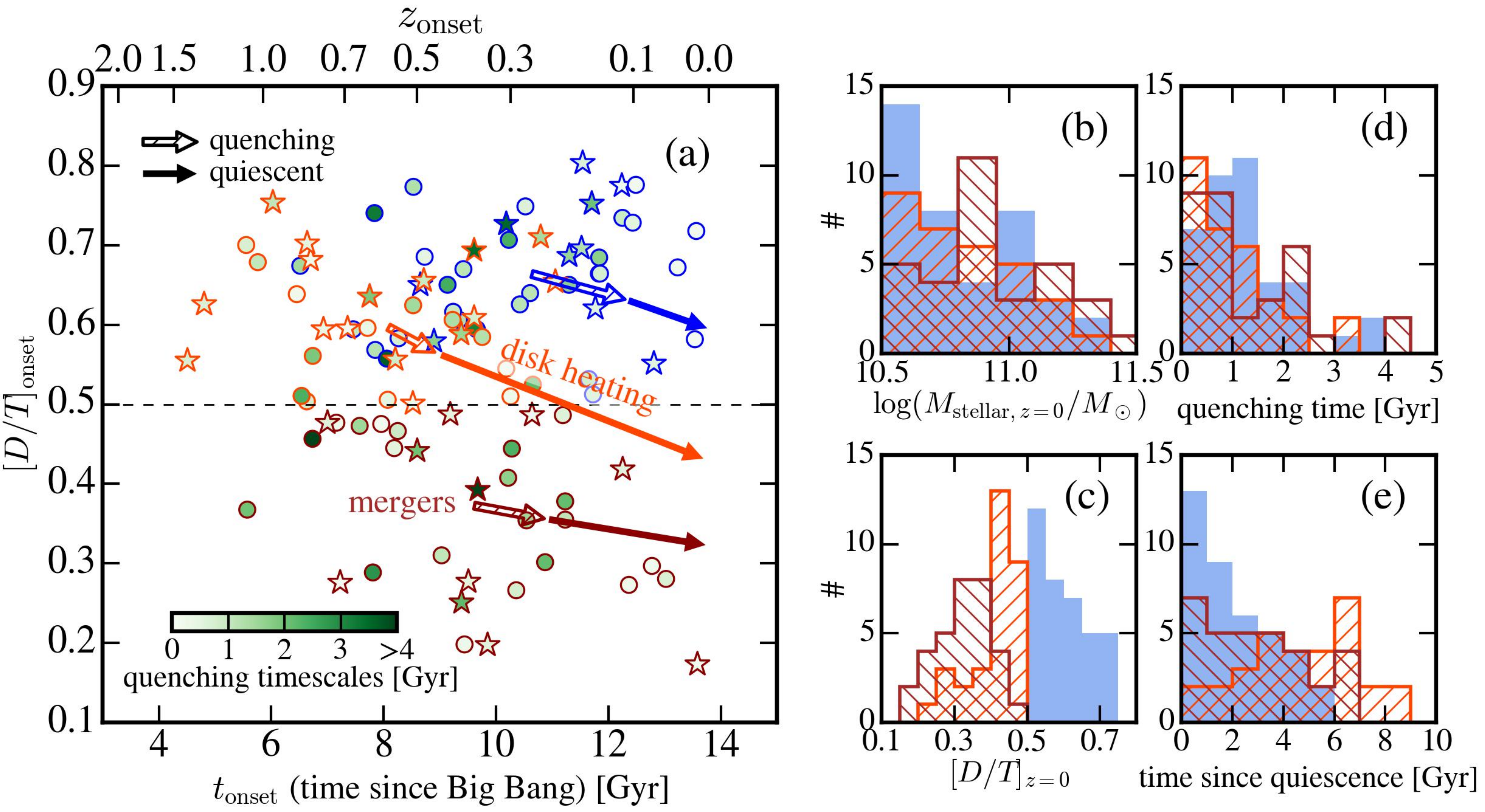}
    \caption{The morphological evolutionary pathways of TNG50 galaxies to quiescent disc and elliptical galaxies at $z=0$. (a) The morphology of galaxies at the onset of quenching ($[D/T]_{\rm onset}$) is plotted at the epoch of the onset of quenching ($t_{\rm onset}$) in the time axis (time since Big Bang in Gyr). Quiescent galaxies are divided into three groups depending on their morphology at $z=0$ and at the onset of quenching: quiescent disc galaxies shown as blue edges, quiescent elliptical galaxies that started to be quenched with discy morphology are plotted as orange edges, and quiescent ellipticals that were already elliptical at the onset of quenching are shown by dark red edges. Each coloured arrow shows the average trajectory in the plane of morphology and time for each group of galaxies during quenching (hatched arrows) and during the quiescent (solid arrows) period. The colour inside the markers indicates the quenching timescales of individual galaxies. Galaxies that are centrals/satellites at the onset of quenching are plotted as circles/stars. The properties of the galaxies in the three groups are shown as histograms in different colours according to the same colour code used in tracks in the left panel: (b) stellar mass at $z=0$, (c) $D/T$ at $z=0$, (d) quenching timescales, and (e) time since quiescence.} 
    \label{fig:DTevolution_for_discs_ellips}
\end{figure*}

In Fig.~\ref{fig:DTevolution_for_discs_ellips}, we have seen that almost half of the quiescent elliptical galaxies (32 galaxies out of 63)  started to be quenched with elliptical morphology. Furthermore, we confirm that only 3 of these 32 galaxies have always been elliptical throughout their history, at least since $z\sim2$. The majority of the galaxies that started to be quenched with elliptical morphology were discy galaxies at earlier times, meaning that they changed their morphology from discs to ellipticals before quenching.

\begin{figure}
    \centering
    \includegraphics[width=\columnwidth]{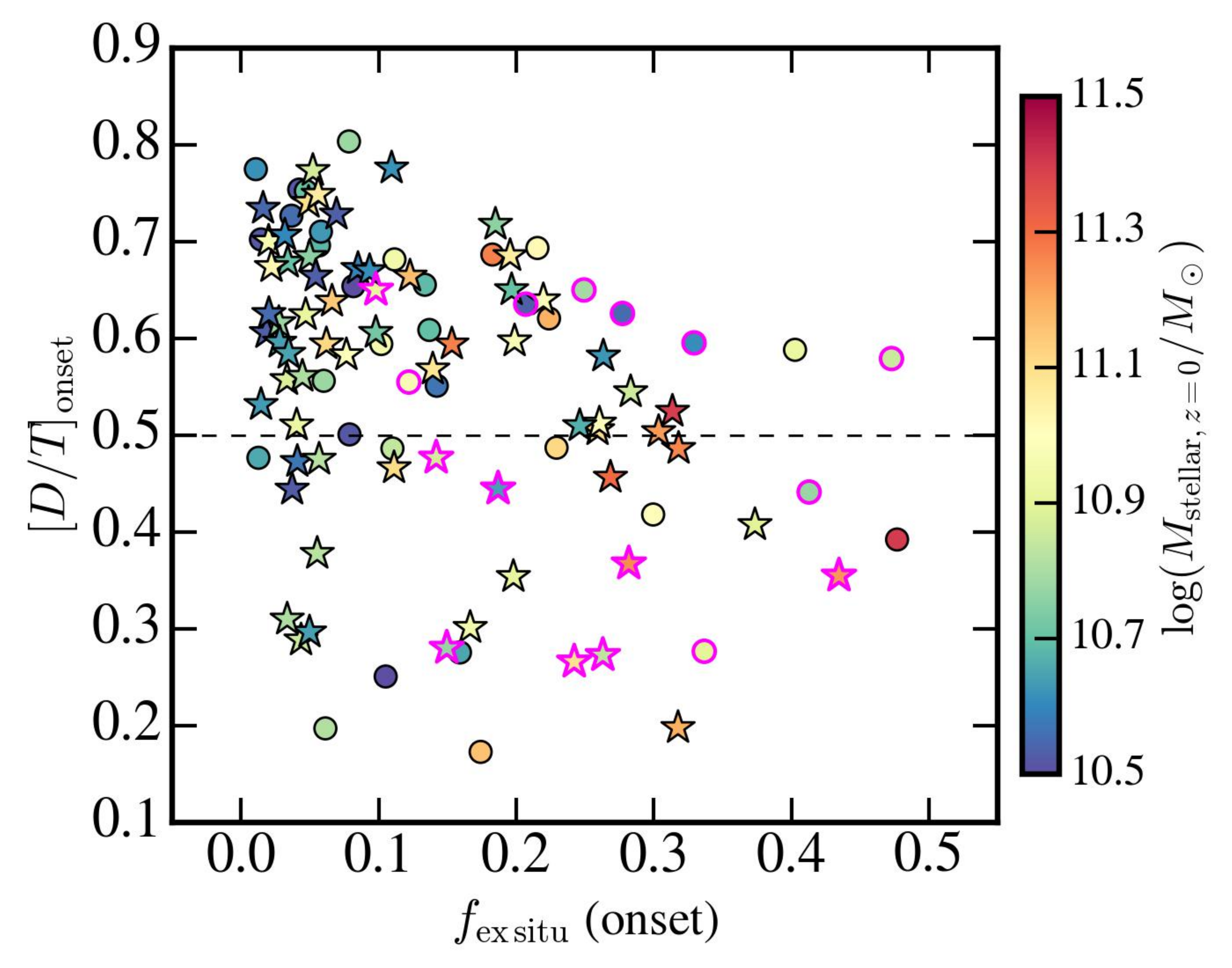}
    \caption{The $D/T$ at the onset of quenching for each TNG50 galaxy as a function of the fractions of stellar particles formed ex situ (\fexsitu) measured at the onset of quenching. Galaxies are colour-coded by final stellar mass at $z=0$, and central and satellite galaxies are shown as circles and stars, respectively.
    The magenta edges show the galaxies that had significant mergers ($>10\%$ of the stellar mass of the galaxy in question) over the last $2\,\rm Gyr$ before the quenching begins.
    %; i.e., the galaxies where the increase in ex-situ mass for the last 2 Gyr divided by galaxy stellar mass 2 Gyr ago is greater than 0.1 ($\Delta M_{ex\,situ}/M_{\rm stellar, 2\,Gyr\,ago}>0.1$). 
    We find that mergers affect the morphology of galaxies at the onset of quenching, but in most cases, they do not seem to immediately shut down the star formation in the galaxies.}
    \label{fig:onset_morp}
\end{figure}

To see what affects the morphology of galaxies at the onset of quenching, Fig.~\ref{fig:onset_morp} shows the $D/T$ at the onset of quenching for each galaxy as a function of the fraction of stellar particles formed ex situ (\fexsitu) at the same epoch (when the quenching begins) as a proxy of the significance of mergers that galaxies had experienced until then. The \fexsitu\, was measured using all the stellar particles in each galaxy following \cite{Rodriguez-Gomez2016TheStars}.
The colour code indicates the stellar mass of the galaxies at $z=0$, and the circles and stars indicate central and satellite galaxies, respectively.

Broadly, galaxies starting to be quenched with elliptical morphology (with low $[D/T]_{\rm onset}$) have higher fractions of stellar particles formed ex situ, meaning that mergers may have turned their morphology to elliptical before quenching begins. As the colour indicates, these galaxies tend to be massive galaxies that are formed as a result of frequent/significant mergers.

However, the mergers that contributed to the elliptical morphology do not seem to trigger the quenching for these galaxies immediately. We identify the galaxies that have significant mergers ($>10\%$ of the stellar mass of the galaxy in question) over the last $2\,\rm Gyr$ before the quenching begins; i.e., the increase in the ex-situ mass for the last $2\,\rm Gyr$ divided by galaxy stellar mass $2\,\rm Gyr$ ago is greater than 0.1 ($\Delta M_{ex\,situ}/M_{\rm stellar, 2\,Gyr\,ago}>0.1$). We mark these galaxies with magenta edges, and there are only a few galaxies where the merger epochs are correlated with the onset of quenching\footnote{If mergers induce starbursts (forming many stars in situ), \fexsitu\, might have not increased so much. Thus, instead of $\Delta$\fexsitu, we used $\Delta M_{ex\,situ}/M_{\rm stellar, 2\,Gyr\,ago}>0.1$ to identify the galaxies where quenching seems to be related to mergers.}. This means that mergers seem to be the mechanism transforming the morphology of galaxies from discy to elliptical but they are not directly responsible for shutting down the star formation in these galaxies. Indeed, some studies using simulations directly measured the epoch of (last major) mergers and the epoch when the quenching begins and found that the two epochs are not tightly correlated \citep[e.g.,][]{Rodriguez-Montero2019, Quai2021, Pathak2021}.
\cite{Weinberger2018} also showed using the TNG simulation that a large population ($>60$ percent) of galaxies are quenched unrelated to a major merger event.

Interestingly, there are some low-mass galaxies (bluer circles/stars in the lower-left corner) that have also transformed their morphologies to elliptical before quenching ($[D/T]_{\rm onset} <0.5$) without experiencing significant mergers (having low \fexsitu\,at the onset of quenching). We visually checked the evolution of these galaxies and found that they change their morphology from discs to ellipticals after having misaligned gas accretion or very gas-rich minor mergers \citep[e.g.,][]{Khim2021}, as they have low contribution in ex-situ mass. Indeed, several studies have shown that counter-rotating streams can lower the angular momentum of gas disc and lead gas to the central regions, building bulges, while galaxies are still on the main sequence \citep[e.g.,][]{Sales2012,Tacchella2016,Tacchella2016Thereplenishment, Park2019, Dekel2020spinflips}. These galaxies would be quenched later, which can explain how some low-mass galaxies have rather elliptical morphology when they start quenching without significant mergers.

\section{Discussion} \label{sec:discussion}
\subsection{The formation of massive quiescent TGN50 galaxies with diverse morphologies}
\label{sec:discuss_variety_of_morphology}

\begin{figure*}
    \centering
    \includegraphics[width=\textwidth]{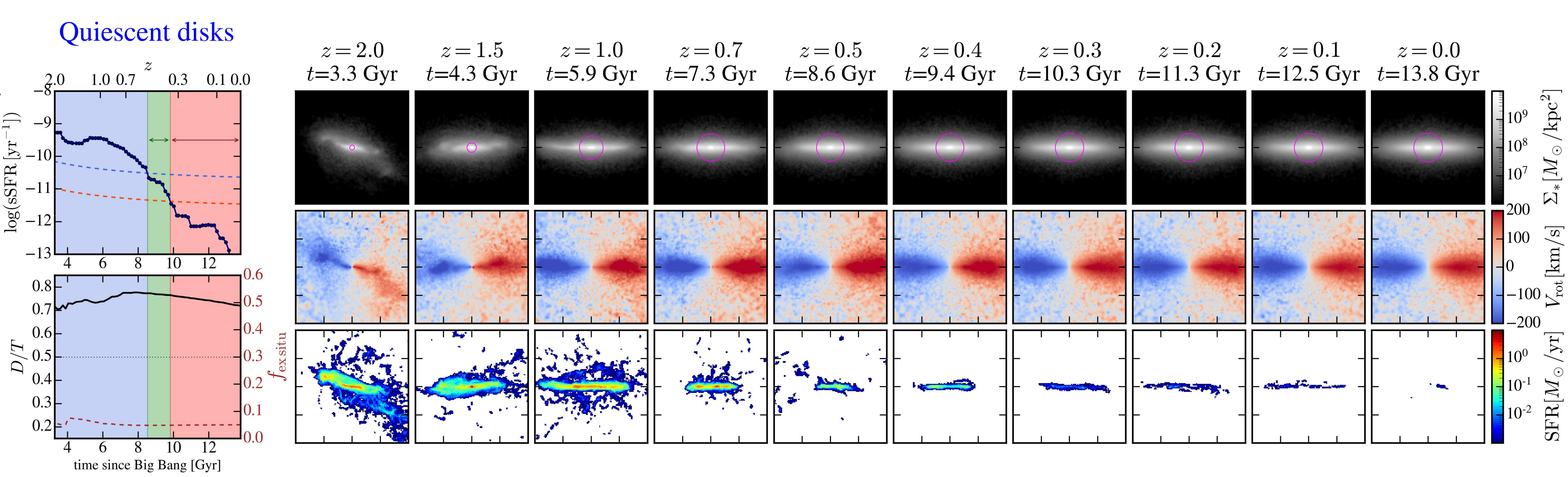}
    \caption{The evolution of one example quiescent disc galaxy at $z=0$. The leftmost panel shows the evolution of the sSFR (top) and $D/T$ (black solid line) and $f_{\rm ex\,situ}$ (brown dashed line). The blue, green, and red regions indicate the star-forming phase, transition phase (thus, quenching timescale), and quiescent phase (time since quiescence) based on the definition using the sSFR criterion (See Section~\ref{sec:methodology}). The right panels show the edge-on images of the galaxy in stellar mass density (top), line-of-sight velocity (middle), and SFR (bottom) from $z=2.0$ to $z=0.0$. The magenta circles in the stellar density maps indicate the half-mass radii. We find that massive quiescent TNG50 discs are recently-quenched disc galaxies, the quenching proceeds slowly in an inside-out manner. Without significant mergers, they can retain their disc structures and are identified as quiescent disc galaxies at $z=0$.  
    }
    \label{fig:quiescent_disc}
\end{figure*}

\begin{figure*}
    \centering
    \includegraphics[width=\textwidth]{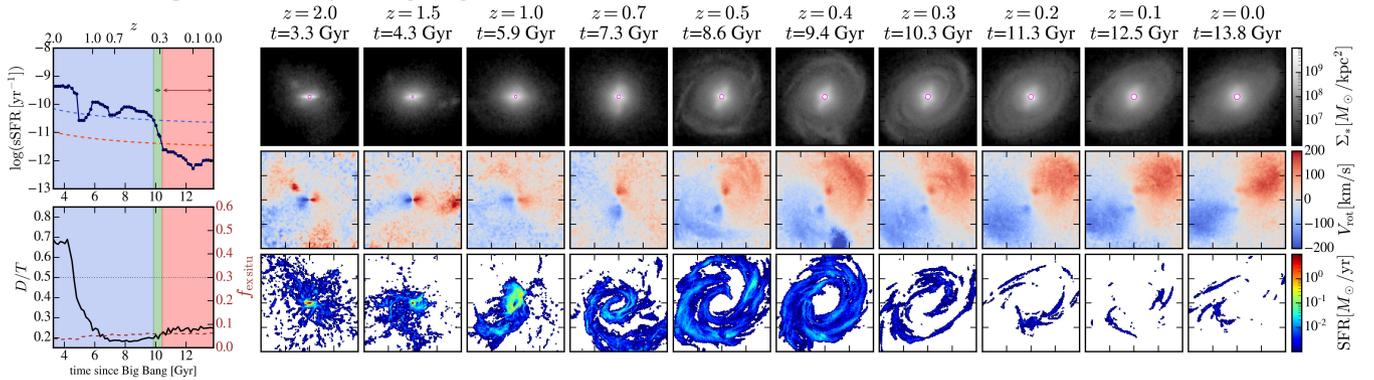}
    \caption{Same format as Fig.~\ref{fig:quiescent_disc} for quiescent elliptical galaxies at $z=0$.  (a) Quiescent ellipticals (that were discs at the onset of quenching), possibly formed by disc heating (following the orange track in Fig.~\ref{fig:DTevolution_for_discs_ellips}). (b)  Quiescent ellipticals that trasformed their morphology to elliptical before quenching by mergers (brown track) and by (c) misaligned gas accretion (or very gas-rich minor mergers).} 
    \label{fig:quiescent_ellipticals}
\end{figure*}

\begin{figure*}
    \centering
    \includegraphics[width=\textwidth]{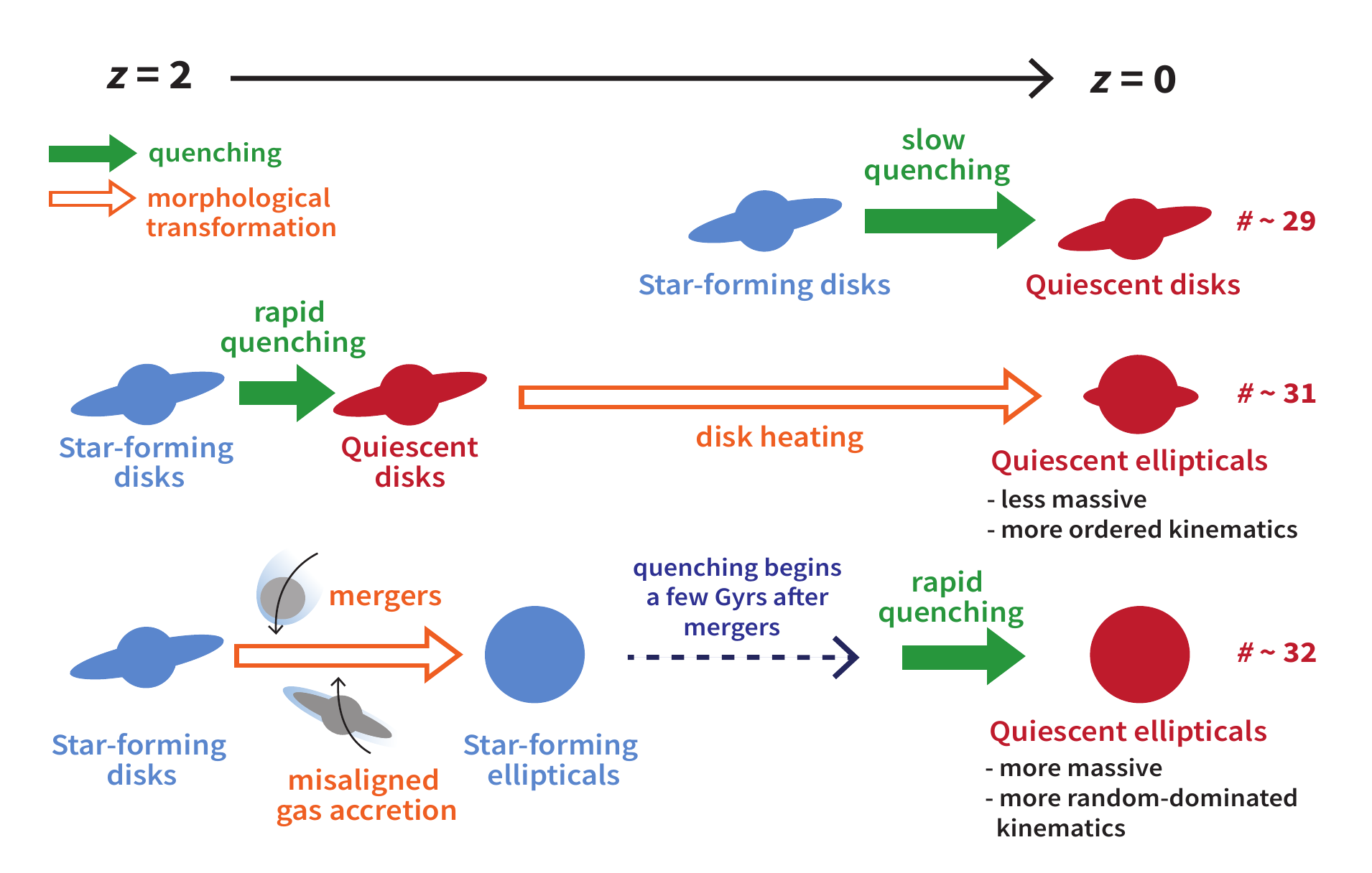}
    \caption{The schematic diagram summarizing the evolutionary pathways towards quiescent disc and elliptical galaxies at $z=0$. The colour of galaxies indicates the star-forming activities of galaxies: star-forming (blue) and quiescent (red). The green and orange arrows indicate the quenching and morphological transformation events. 
    We find that massive quiescent TNG50 discs are recently and slowly quenched disc galaxies. For massive quiescent ellipticals in TNG50, half (31/63) of them are formed as a result of quiescent discs being heated for a long time by internal and external processes, as the example shown in Fig.~\ref{fig:quiescent_ellipticals}(a). The other half (32/63) transform their morphology before quenching primarily by mergers, as shown in Fig.~\ref{fig:quiescent_ellipticals}(b). However, the mergers that contributed to morphological transformation do not immediately quench galaxies in many cases. In some cases, galaxies become elliptical/spheroidal before quenching by misaligned gas accretion (without significant mergers), as shown in Fig.~\ref{fig:quiescent_ellipticals}(c).}
    \label{fig:summary}
\end{figure*}

As described in the introduction, several possible quenching mechanisms including both internal and external processes have been suggested and explored for a long time. While the importance of each mechanism seems to depend on galaxy stellar mass, environment, and model, it appears that AGN feedback is the key driver of the quenching of massive central galaxies in the TNG model in the mass regime above $\log(M_{\rm stellar}/M_{\odot})\sim10.5$ \citep[e.g.,][]{Weinberger2018, Nelson2018FirstBimodality}. For satellites, while we do see some hints of the environmental effects on quenching satellite galaxies in the massive groups and clusters on short timescales, the majority of the massive quiescent TNG50 galaxies are central galaxies. 
Moreover, \cite{Donnari2021a} have shown with TNG100, where the volume includes diverse environments, that massive galaxies (above a few times $10^{10}\,\msun$) are more likely to be quenched by AGN feedback regardless of the environment; i.e., regardless of whether they are centrals or satellites.

With seemingly the same dominant quenching mechanism -- AGN feedback --, however, galaxies end up having a variety of morphologies depending on the degree of AGN feedback and the stochastic merger history. As we showed in Fig.~\ref{fig:quenching_quiescent_time_DT}, quiescent disc galaxies are mostly recently quenched and they are quenched more slowly. As Fig.~\ref{fig:onset_props_quenching_time}(c) and (d) indicates, longer quenching timescales seem to be associated with lower (but still strong enough to quench) AGN feedback energy. Many observational studies \citep[e.g.,][]{Tacchella2015Sci, Nelson2016, Ellison2018, Tacchella2018, Morselli2019} have suggested that quenching proceeds in an inside-out fashion; central regions are quenched first while disc regions in the outskirt continue to form new stars and later quenched as well. Recently, \cite{Nelson2021} have studied the evolution of spatially resolved star formation using the TNG50 galaxies and showed that TNG50 galaxies are also quenched inside-out.

Fig.~\ref{fig:quiescent_disc} shows the evolution of one example quiescent disc galaxy at $z=0$.
The leftmost top panel shows the evolution of sSFR and the bottom panel shows the evolution of $D/T$ (black solid line) and $f_{\rm ex\,situ}$ (brown dashed line, on the right y-axis). The blue, green, and red regions are the star-forming, transition, and quiescent phase, as defined using the sSFR (See Section~\ref{sec:methodology}); thus, the duration of the green and red regions correspond to the quenching timescale and time since quiescence. The right panels show the edge-on images of the galaxy in stellar mass density (top), line-of-sight velocity (middle), and SFR (bottom) from $z=2.0$ to $z=0.0$. 
As the bottom panel shows the SFR of this galaxy is, indeed, quenched first in the central regions, while there is still residual star formation in the outskirts (disc regions).

Therefore, the massive quiescent discs in TNG50 are recently quenched disc galaxies with mild AGN feedback.
While disc galaxies are quenched rather slowly and in an inside-out manner, and without significant mergers, they can retain their disc structures and are identified as quiescent disc galaxies at $z=0$.

Disc galaxies that are quenched very rapidly at higher redshifts with strong AGN feedback could form quiescent ellipticals at $z=0$ (see the orange track in Fig.~\ref{fig:DTevolution_for_discs_ellips}). In this case, morphological transformation occurs on a long timescale after quenching. 
Among those 31 galaxies that were discs at the onset of quenching and decrease their $D/T$ while spending a long time being quiescent, only 3 of them experience significant mergers after quenching (the change in ex-situ mass fraction, $\Delta f_{\rm ex\,situ}>0.1$ during quiescence). 
The majority, remaining galaxies gradually decrease their $D/T$ until $z=0$, possibly by disc heating. 
Here, by disc heating, we broadly refer to the change in the orbital properties of stellar particles, which can be driven by both internal secular processes like bar heating \citep[e.g.,][]{Grand2016VerticalContext} and external perturbations such as fly-bys or tidal shock heating at pericentre in the case of cluster satellites \citep[see][]{Joshi2020TheClusters}. At $z=0$, they end up being dominated by stellar particles with random motions (thus, $D/T<0.5$). Although they are classified as ellipticals based on their kinematics, most retain some degree of rotation, with $D/T\sim 0.3-0.5$ (orange hatched histogram in Fig.~\ref{fig:DTevolution_for_discs_ellips}(c)).

Fig.~\ref{fig:quiescent_ellipticals} (a) shows the evolution of one quiescent elliptical galaxy at $z=0$ that is thought to be formed by disc heating. This galaxy is very rapidly quenched at $z\sim0.8$. Quiescent since then, its disc is gradually heated without significant/major mergers (low $f_{\rm ex\,situ}$ throughout the time, as shown as the brown dashed line in the leftmost bottom panel). Eventually, this galaxy becomes thicker/spheroidal (kinematically dominated by random motions, lower $D/T$) to be identified as a quiescent elliptical at $z=0$.

Quiescent ellipticals with the most random motion-dominated kinematics (the brown track in Fig.~\ref{fig:DTevolution_for_discs_ellips}) are formed by disc galaxies at higher redshifts building up large spheroidal components by significant mergers before the onset of quenching. These significant mergers transform the bulk kinematics of galaxies from disc-dominated to random-motion-dominated (with $D/T\sim0.3$, as shown as brown hatched histogram in Fig.~\ref{fig:DTevolution_for_discs_ellips}(c)) and also increase the stellar mass of galaxies; they are overall more massive than the quiescent ellipticals quenched at higher redshifts (shown by the orange hatched histogram). This is consistent with the findings of \cite{Du2021}: they studied the evolution of kinematically-decomposed structures, e.g., discs, bulges, and halos, and found that massive bulge-dominated galaxies develop compact massive bulges at high-$z$ as a result of mergers, and in the later phase, these galaxies are quenched inside-out.

However, quenching in these galaxies does not occur immediately after the mergers (Fig.~\ref{fig:onset_morp}).
This has been found in other previous studies using simulations \citep[e.g.,][]{Rodriguez-Montero2019, Quai2021, Pathak2021}.
Perhaps, it takes some time for the SMBHs to grow by mergers and release enough feedback energy to quench galaxies. 
Also, mergers are more likely to be gas-rich at higher redshifts, thus, the merger remnant would be actively forming stars immediately after merging and it may take some time until the galaxies consume all the gas and start to be quenched. 
This is consistent with the results of \cite{Zolotov2015CompactionNuggets} and \cite{Tacchella2016}: merger-like events (including misaligned gas accretion) could induce ``compaction'' and the formation of spheroidal components in star-forming galaxies, but quenching happens later and does not need to be directly associated with the events \citep[see also][]{Dekel2014WetNuggets, Dekel2019arXiv}
This may explain why there are not many galaxies that started to be quenched as ellipticals at higher redshifts (i.e., not many galaxies in the lower left corner in Fig.~\ref{fig:DTevolution_for_discs_ellips}(a)) and why these quenched ellipticals started to appear at later epochs. In other words, star-forming blue ellipticals/spheroids would be more frequently found at higher redshifts.

Indeed, Fig~\ref{fig:quiescent_ellipticals} (b) shows one example of a quiescent elliptical galaxy formed by mergers. This galaxy has a significant merger at $t\sim 7\,\rm Gyr$ (the $f_{\rm ex\,situ}$ soars to $\sim 0.4$); as a result, the morphology becomes highly disturbed with low $D/T$. Even after the morphology is transformed to elliptical/spheroidal, the galaxy is still star-forming, and it starts to be quenched $\sim 3\,\rm Gyr$ after the merger (at $t\sim10\,\rm Gyr$). Thus, this example clearly shows that the mergers that contributed to morphological transformation do not immediately quench galaxies.

Also, some elliptical galaxies develop large spheroidal components before quenching by misaligned gas accretion (without significant mergers), one of which is shown in Fig.~\ref{fig:quiescent_ellipticals} (c). As stars are still formed from gas accreted in a misaligned direction with the net galactic rotation, the bulk kinematics of the galaxy remains random-motion-dominated (low $D/T$ on the bottom left panel). The galaxy starts to be quenched at a much later epoch after it consumes all the gas.

Fig.~\ref{fig:summary} is a schematic diagram summarizing the evolutionary pathways towards quiescent disc and elliptical galaxies.
In conclusion, we find that massive quiescent TNG50 discs are recently quenched disc galaxies, and they could retain their disc structures as the quenching proceeds slowly by mild AGN feedback in an inside-out manner and there are no significant mergers that could disrupt their discs. 
On the other hand, half of the quiescent elliptical galaxies at $z=0$ are formed mainly as a result of long-term heating of disc galaxies at higher redshifts that are rapidly quenched by strong AGN feedback. 
These quiescent ellipticals are slightly less massive and retain some degree of rotation at $z=0$.
The other half transform their morphology before quenching primarily by merger, or in some cases, by misaligned gas accretion (or very gas-rich minor mergers), both of which could induce strong AGN feedback responsible for rapid quenching.
These quiescent ellipticals are more massive and have the most random-dominated kinematics.

\subsection{Caveats} \label{sec:discuss_caveats}
Here we discuss some caveats of this study and show how our results might change both qualitatively and quantitatively if we try different criteria for some of our measurements.

\subsubsection{Numerical heating}
It is worth noting that simulated galaxies might suffer from numerical (artificial) heating caused by interactions between the stellar and dark matter particles. This could artificially heat the stellar discs of galaxies and make them look thicker and rounder, thus eventually altering galactic morphology. The effect of numerical heating has been explored in many studies. For example, recently, \cite{Ludlow2021} have run a suite of simulations of stellar discs embedded in a dark matter halo resolved with dark matter particles of different masses and concluded that the degree of numerical heating is determined by the DM particle mass. As the TNG50 simulation we used for this study has one of the highest DM particle resolutions among the cosmological volume simulations ($m_{\rm DM}\sim4.5\times10^5\,\msun$), we can argue that our simulated galaxies would be less affected by numerical heating. Moreover, \cite{Joshi2020TheClusters} have clearly shown that numerical heating is not an issue in the TNG50 galaxies in terms of making galaxies more spheroidal; they have found that the majority of TNG50 disc galaxies that do not undergo environmental effects (e.g., mergers, fly-bys, and tidal shocking at pericentric passages) do not become less discy with time.

\subsubsection{Aperture size}
In this study, we used $R_{\rm full}$, defined as the radius where the surface stellar mass density drops below $\Sigma=10^6\,\msun\rm /kpc^2$, as an aperture size within which the SFR, stellar mass, and the morphology indicator ($D/T$) are measured. While this $R_{\rm full}$ attempts to mimic the surface brightness cuts done for many local galaxies, covering most of the visual extent of galaxies, it might be difficult to perform on galaxies at higher redshifts. Typically, many of the properties for high-$z$ galaxies are measured within a few times the effective radii, e.g., $1-3\,R_{\rm eff}$, which sometimes covers only a fraction of the entire extent of the galaxies. So, to see how our results would change if we used different aperture sizes, we repeated our analysis again using $3\,R_{\rm eff}$ as the aperture for measuring stellar mass, star formation rate, and $D/T$. On average, the aperture of $3\,R_{\rm eff}$ is 3 times smaller than $R_{\rm full}$.

For the most part, our results remain unchanged even if we use $3\,R_{\rm eff}$ instead of $R_{\rm full}$. The number of quiescent galaxies selected at $z=0$ is slightly increased from 102 to 131, suggesting that there are some galaxies where the central regions are totally quenched (thus, classified as quiescent galaxies with $3\,R_{\rm eff}$) while there is residual star formation in the outskirts/disc regions (thus, identified as non-quiescent with $R_{\rm full}$). When using $3\,R_{\rm eff}$ (which is $\sim 3$ times smaller than $R_{\rm full}$), one of the most noticeable changes is that the majority of quiescent galaxies (102 out of 131) are classified as ellipticals ($D/T<0.5$). This means that many of the quiescent discs we originally selected have highly concentrated mass in the centre of galaxies so that they are classified as ellipticals with smaller apertures, but in fact they do have (fainter) disc structures (rotating component) in the outskirts.

The trends of quenching history for disc and elliptical galaxies measured with $3\,R_{\rm eff}$ are roughly similar to our results as shown in Fig.~\ref{fig:quenching_quiescent_time_DT}; quiescent discs are mostly recently quenched and many of them are quenched slowly, whereas ellipticals are mostly quenched rapidly and have a wide range of time since quiescence. With reduced aperture, a higher fraction of elliptical galaxies are quenched on short timescales; with $R_{\rm full}$, 32\% of ellipticals are quenched within $<0.5\,\rm Gyr$, and this fraction is increased to $42\%$ with $3\,R_{\rm eff}$. This suggests that central concentrated regions are quenched more rapidly than the outer regions, consistent with an inside-out quenching scenario.

We also confirm that our qualitative conclusion on the formation of quiescent galaxies with a variety of morphologies discussed in the previous section remains still valid with different choices of aperture such as $3\,R_{\rm eff}$. Though, the fraction of elliptical galaxies that change their morphology after they are quenched (equivalent to the orange track in Fig.~\ref{fig:DTevolution_for_discs_ellips}) has become slightly increased from a half to $\sim 60\%$ when a smaller aperture is used; the remaining $40\%$ of ellipticals changed their morphology prior to the onset of quenching by mergers.

\subsubsection{The choice of $D/T$ cut}
The morphology of galaxies is quantified based on the mass fractions of the kinematically selected disc stars ($\epsilon>0.5$). Also, we distinguish between disc and elliptical galaxies by the arbitrary cut of $D/T=0.5$. In Section~\ref{sec:evolutionary_parths}, we examined the average evolutionary track to $z=0$ quiescent disc as well as quiescent ellipticals (see Fig.~\ref{fig:DTevolution_for_discs_ellips}(a)) and found that approximately half of the quiescent ellipticals are the rapidly quenched disc galaxies at high redshifts that transformed their morphology from discy to elliptical, while the other half changed their morphology much earlier than the epoch when the quenching begins.
This quantitative result might depend on the choice of the cut. For example, if we lower our $D/T$ cut as $D/T=0.35$, the majority of quiescent ellipticals transformed their morphology prior to the onset of quenching.

However, with the lowered (higher) $D/T$ cut, there would be more quiescent discs (quiescent ellipticals) at $z=0$ which does not agree with the observed morphological mix. 
Therefore, while we admit that our quantitative results depend on the choice of $D/T$ cut, the choice of $D/T=0.5$ is the most reasonable cut that reproduces the morphological mix in observations. 
Qualitatively, the fact that quiescent disc galaxies are formed as recently-quenched disc galaxies and that many of the quiescent ellipticals transformed their morphology by mergers prior to the quenching would not change.

\subsubsection{Cluster environments}
%While most of the quiescent galaxies in the local Universe reside in dense environments, including within massive groups and clusters, the TNG50 simulation that we use in this study has only two Virgo-like clusters with virial masses of $M_{\rm 200\,c}\sim10^{14}\,\msun$. 
%While some of our massive quiescent galaxies are cluster satellites with host halo masses of $\log(M_{\rm 200\,c}/M_\odot)\sim 13.5-14.3$, we have not studied the quenching history and morphological evolution of these cluster galaxies in detail. 

%Therefore, the quiescent galaxies that we study might not be a fully representative sample of local massive quiescent galaxies in the massive clusters. 
In TNG50, there are two Virgo-like clusters with virial masses of $M_{\rm 200\,c}\sim10^{14}\,\msun$. Indeed, 30 of 102 quiescent galaxies in our sample are selected as cluster satellites with host halo masses of $\log(M_{\rm 200\,c}/M_\odot)\sim 13.5-14.3$, and they are mostly rapidly quenched ($<1\,\rm Gyr$, in Fig.~\ref{fig:onset_props_quenching_time}(a)) possibly due to environmental effects. 
In this work, we have not studied in detail how galaxies are quenched and transform their morphology as they fall into clusters \citep[see ][for more details]{Joshi2020TheClusters}. We also do not have satellites of massive clusters, as massive as Coma ($\sim 10^{15}\,\msun$), where environmental effects are believed to be more dramatic. 
In much denser environments, galaxies would experience mergers more often in their early formation histories  \citep[e.g.,][]{Muldrew2018}, which would likely make them more spheroidal. As they would become more massive and host larger SMBHs, many of them would start to be quenched with their elliptical morphology just as the brown track shown in Fig.~\ref{fig:DTevolution_for_discs_ellips}. 

%The evolution of ``cluster'' quiescent galaxies is not well studied in this work.
%it would be worthwhile to look into more detail how galaxies are quenched and transform their morphology as they fall into clusters \citep[e.g.,][]{Joshi2020TheClusters}, eventually to understand the density-morphology relation \citep[e.g.,][]{Dressler1980GalaxyGalaxies}. 

\section{Summary and Conclusion} \label{sec:summary}
In this study, we use massive quiescent TNG50 galaxies with stellar mass in a range of $10.5<\log(M_{\rm stellar}/M_{\odot})<11.5$, including both centrals and satellites, to explore how quiescent galaxies end up having various morphologies. Specifically, we aim to answer these questions: (i) how does the quenching timescale differ for galaxies with different morphologies? (ii) what affects the quenching timescales? and (iii) when does  morphological transformation happens: before, during, or after quenching?

We select quiescent galaxies at $z=0$ as the galaxies with ${\rm sSFR} <1/[20\,t_{\rm H}(z)]$, and quantify the morphology of galaxies using the kinematic morphological indicator, $D/T$ (the mass fraction of stellar particles rotating in near-circular orbits as disc stars). As a result, we have in total 102 quiescent galaxies $z=0$ consisting of 39 disc ($D/T>0.5$) and 63 elliptical ($D/T<0.5$) galaxies. We directly track the evolution in sSFR of individual galaxies and measure the quenching timescale (i.e., how fast the sSFR drops) and the time since quiescence (i.e., how long galaxies have been quiescent). Here we summarize our key findings.

\begin{enumerate}
    \item We find that quiescent galaxies in the TNG50 simulation have a variety of quenching timescales and time since quiescence, which varies depending on the morphology of galaxies; quiescent discs are mostly recently and slowly quenched, whereas quiescent ellipticals have a wide range of time since quiescence, and most of them are rapidly quenched. 
    \item We explored what affects the quenching timescales and find that galaxies are quenched more rapidly if they: (i) are satellites in massive halos, (ii) have lower star-forming gas fractions, or (iii) inject a larger amount of kinetic AGN feedback energy. Rejuvenated galaxies are not common in the TNG50 simulation and they tend to have longer quenching timescales.  
    \item Quiescent discs in the TNG50 result from recently-quenched disc-like galaxies where quenching proceeds slowly (possibly by relatively milder AGN feedback) in an in-side out manner. Quiescent ellipticals are formed by two paths.
    (i) Less massive and kinematically colder ellipticals are formed by disc galaxies that are quenched rapidly by strong AGN feedback at higher redshifts and transform their morphology to ellipticals while being quiescent for a long time. (ii) More massive and kinematically hotter ellipticals are formed by galaxies building up large spheroidal components as a result of mergers prior to quenching. However, the mergers that contributed to the morphological transformation do not always immediately quench galaxies in most cases.
\end{enumerate}

In conclusion, we find that {\em quenching and the morphological transformation are decoupled}; half of the ellipticals gradually transformed their morphology {\em after} quenching, while the other half transformed {\em before} quenching by mergers.
We conclude that AGN feedback and the stochastic merger history leads to a diversity of massive quiescent galaxies in the field environments in terms of both morphology and quenching timescales. 
Future investigations can further study the process of SMBH accretion and growth in massive galaxies, as well as the detailed relationship between multi-scale gas inflows and galaxy quenching.
%What feeds the SMBHs in massive galaxies and the detailed sequence in which the BHs grow and AGN feedback energy is released to quench galaxies is beyond the scope of our work, but is worth investigating further.
Future high-resolution simulations in much larger volumes including several massive clusters will provide novel insights on the quenching of massive quiescent galaxies. 
Finally, we expect that future observations such as James Webb Space Telescope (JWST) could provide more information about the morphology of high-$z$ galaxies and when and how quiescent galaxies first emerged in cosmic history. 

\section*{Data Availability}
Data directly related to this publication and its figures is available on request from the corresponding author. 
The IllustrisTNG simulations, including TNG50 used in this study, are publicly available and accessible at www.tng-project.org/data \cite{Nelson2019}. 

\section*{Acknowledgements}
The primary TNG simulations including TNG50 were carried out with compute time granted by the Gauss Centre for Supercomputing (GCS) under Large-Scale Projects GCS-ILLU and GCS-DWAR on the GCS share of the supercomputer Hazel Hen at the High Performance Computing Center Stuttgart (HLRS).
DN acknowledges funding from the Deutsche Forschungsgemeinschaft (DFG) through an Emmy Noether Research Group (grant number NE 2441/1-1).
RW was supported by the Natural Sciences and Engineering Research Council of Canada (NSERC), funding reference $\#$ CITA 490888-16. 
S.T. is supported by the 2021 Research Fund 1.210134.01 of UNIST (Ulsan National Institute of Science $\&$ Technology).
The computations in this paper were run on the
FASRC Cannon cluster supported by the FAS Division of Science Research Computing Group at Harvard University. 
This research used the Python packages \texttt{Matplotlib} \citep{Hunter2007}, \texttt{Numpy} \citep{Harris2020NumPy-Array}, and \texttt{Colossus} \citep{Diemer2018ApJS}.

%%%%%%%%%%%%%%%%%%%%%%%%%%%%%%%%%%%%%%%%%%%%%%%%%%



\bibliographystyle{mnras}
\bibliography{references} % if your bibtex file is called example.bib

\begin{thebibliography}{}
\makeatletter
\relax
\def\mn@urlcharsother{\let\do\@makeother \do\$\do\&\do\#\do\^\do\_\do\%\do\~}
\def\mn@doi{\begingroup\mn@urlcharsother \@ifnextchar [ {\mn@doi@}
  {\mn@doi@[]}}
\def\mn@doi@[#1]#2{\def\@tempa{#1}\ifx\@tempa\@empty \href
  {http://dx.doi.org/#2} {doi:#2}\else \href {http://dx.doi.org/#2} {#1}\fi
  \endgroup}
\def\mn@eprint#1#2{\mn@eprint@#1:#2::\@nil}
\def\mn@eprint@arXiv#1{\href {http://arxiv.org/abs/#1} {{\tt arXiv:#1}}}
\def\mn@eprint@dblp#1{\href {http://dblp.uni-trier.de/rec/bibtex/#1.xml}
  {dblp:#1}}
\def\mn@eprint@#1:#2:#3:#4\@nil{\def\@tempa {#1}\def\@tempb {#2}\def\@tempc
  {#3}\ifx \@tempc \@empty \let \@tempc \@tempb \let \@tempb \@tempa \fi \ifx
  \@tempb \@empty \def\@tempb {arXiv}\fi \@ifundefined
  {mn@eprint@\@tempb}{\@tempb:\@tempc}{\expandafter \expandafter \csname
  mn@eprint@\@tempb\endcsname \expandafter{\@tempc}}}

\bibitem[\protect\citeauthoryear{Abadi, Navarro, Steinmetz  \& Eke}{Abadi
  et~al.}{2003}]{Abadi2003SimulationsDisks}
Abadi M.~G.,  Navarro J.~F.,  Steinmetz M.,   Eke V.~R.,  2003, \mn@doi [\apj]
  {10.1086/378316}, 597, 21

\bibitem[\protect\citeauthoryear{{Baldry}, {Glazebrook}, {Brinkmann},
  {Ivezi{\'c}}, {Lupton}, {Nichol}  \& {Szalay}}{{Baldry}
  et~al.}{2004}]{Baldry2004}
{Baldry} I.~K.,  {Glazebrook} K.,  {Brinkmann} J.,  {Ivezi{\'c}} {\v{Z}}.,
  {Lupton} R.~H.,  {Nichol} R.~C.,   {Szalay} A.~S.,  2004, \mn@doi [\apj]
  {10.1086/380092}, \href
  {https://ui.adsabs.harvard.edu/abs/2004ApJ...600..681B} {600, 681}

\bibitem[\protect\citeauthoryear{{Barnes}}{{Barnes}}{1988}]{Barnes1988ApJ331}
{Barnes} J.~E.,  1988, \mn@doi [\apj] {10.1086/166593}, \href
  {https://ui.adsabs.harvard.edu/abs/1988ApJ...331..699B} {331, 699}

\bibitem[\protect\citeauthoryear{{Behroozi}, {Wechsler}, {Hearin}  \&
  {Conroy}}{{Behroozi} et~al.}{2019}]{Behroozi2019}
{Behroozi} P.,  {Wechsler} R.~H.,  {Hearin} A.~P.,   {Conroy} C.,  2019,
  \mn@doi [\mnras] {10.1093/mnras/stz1182}, \href
  {https://ui.adsabs.harvard.edu/abs/2019MNRAS.488.3143B} {488, 3143}

\bibitem[\protect\citeauthoryear{{Belli}, {Newman}  \& {Ellis}}{{Belli}
  et~al.}{2019}]{Belli2019}
{Belli} S.,  {Newman} A.~B.,   {Ellis} R.~S.,  2019, \mn@doi [\apj]
  {10.3847/1538-4357/ab07af}, \href
  {https://ui.adsabs.harvard.edu/abs/2019ApJ...874...17B} {874, 17}

\bibitem[\protect\citeauthoryear{{Birnboim} \& {Dekel}}{{Birnboim} \&
  {Dekel}}{2003}]{Birnboim2003}
{Birnboim} Y.,  {Dekel} A.,  2003, \mn@doi [\mnras]
  {10.1046/j.1365-8711.2003.06955.x}, \href
  {https://ui.adsabs.harvard.edu/abs/2003MNRAS.345..349B} {345, 349}

\bibitem[\protect\citeauthoryear{{Bluck}, {Mendel}, {Ellison}, {Moreno},
  {Simard}, {Patton}  \& {Starkenburg}}{{Bluck} et~al.}{2014}]{Bluck2014}
{Bluck} A. F.~L.,  {Mendel} J.~T.,  {Ellison} S.~L.,  {Moreno} J.,  {Simard}
  L.,  {Patton} D.~R.,   {Starkenburg} E.,  2014, \mn@doi [\mnras]
  {10.1093/mnras/stu594}, \href
  {https://ui.adsabs.harvard.edu/abs/2014MNRAS.441..599B} {441, 599}

\bibitem[\protect\citeauthoryear{{Bundy} et~al.,}{{Bundy}
  et~al.}{2010}]{Bundy2010TheCOSMOS}
{Bundy} K.,  et~al., 2010, \mn@doi [\apj] {10.1088/0004-637X/719/2/1969}, \href
  {https://ui.adsabs.harvard.edu/abs/2010ApJ...719.1969B} {719, 1969}

\bibitem[\protect\citeauthoryear{{Chabrier}}{{Chabrier}}{2003}]{Chabrier2003GalacticFunction}
{Chabrier} G.,  2003, \mn@doi [\pasp] {10.1086/376392}, \href
  {https://ui.adsabs.harvard.edu/abs/2003PASP..115..763C} {115, 763}

\bibitem[\protect\citeauthoryear{{Cheung} et~al.,}{{Cheung}
  et~al.}{2012}]{Cheung2012}
{Cheung} E.,  et~al., 2012, \mn@doi [\apj] {10.1088/0004-637X/760/2/131}, \href
  {https://ui.adsabs.harvard.edu/abs/2012ApJ...760..131C} {760, 131}

\bibitem[\protect\citeauthoryear{{Correa}, {Schaye}  \& {Trayford}}{{Correa}
  et~al.}{2019}]{Correa2019}
{Correa} C.~A.,  {Schaye} J.,   {Trayford} J.~W.,  2019, \mn@doi [\mnras]
  {10.1093/mnras/stz295}, \href
  {https://ui.adsabs.harvard.edu/abs/2019MNRAS.484.4401C} {484, 4401}

\bibitem[\protect\citeauthoryear{{Croton} et~al.,}{{Croton}
  et~al.}{2006}]{Croton2006}
{Croton} D.~J.,  et~al., 2006, \mn@doi [\mnras]
  {10.1111/j.1365-2966.2005.09675.x}, \href
  {https://ui.adsabs.harvard.edu/abs/2006MNRAS.365...11C} {365, 11}

\bibitem[\protect\citeauthoryear{{Davies}, {Crain}  \& {Pontzen}}{{Davies}
  et~al.}{2021}]{Davies2021}
{Davies} J.~J.,  {Crain} R.~A.,   {Pontzen} A.,  2021, \mn@doi [\mnras]
  {10.1093/mnras/staa3643}, \href
  {https://ui.adsabs.harvard.edu/abs/2021MNRAS.501..236D} {501, 236}

\bibitem[\protect\citeauthoryear{Dekel \& Burkert}{Dekel \&
  Burkert}{2014}]{Dekel2014WetNuggets}
Dekel A.,  Burkert A.,  2014, \mn@doi [\mnras] {10.1093/mnras/stt2331}, 438,
  1870

\bibitem[\protect\citeauthoryear{Dekel et~al.,}{Dekel
  et~al.}{2009a}]{Dekel2009ColdFormation}
Dekel A.,  et~al., 2009a, \mn@doi [Nature] {10.1038/nature07648}, 457, 451

\bibitem[\protect\citeauthoryear{Dekel, Sari  \& Ceverino}{Dekel
  et~al.}{2009b}]{Dekel2009FormationSpheroids}
Dekel A.,  Sari R.,   Ceverino D.,  2009b, \mn@doi [\apj]
  {10.1088/0004-637X/703/1/785}, 703, 785

\bibitem[\protect\citeauthoryear{{Dekel}, {Lapiner}  \& {Dubois}}{{Dekel}
  et~al.}{2019}]{Dekel2019arXiv}
{Dekel} A.,  {Lapiner} S.,   {Dubois} Y.,  2019, arXiv e-prints, \href
  {https://ui.adsabs.harvard.edu/abs/2019arXiv190408431D} {p. arXiv:1904.08431}

\bibitem[\protect\citeauthoryear{{Dekel}, {Ginzburg}, {Jiang}, {Freundlich},
  {Lapiner}, {Ceverino}  \& {Primack}}{{Dekel}
  et~al.}{2020}]{Dekel2020spinflips}
{Dekel} A.,  {Ginzburg} O.,  {Jiang} F.,  {Freundlich} J.,  {Lapiner} S.,
  {Ceverino} D.,   {Primack} J.,  2020, \mn@doi [\mnras]
  {10.1093/mnras/staa470}, \href
  {https://ui.adsabs.harvard.edu/abs/2020MNRAS.493.4126D} {493, 4126}

\bibitem[\protect\citeauthoryear{{Di Matteo}, {Springel}  \& {Hernquist}}{{Di
  Matteo} et~al.}{2005}]{DiMatteo2005}
{Di Matteo} T.,  {Springel} V.,   {Hernquist} L.,  2005, \mn@doi [\nat]
  {10.1038/nature03335}, \href
  {https://ui.adsabs.harvard.edu/abs/2005Natur.433..604D} {433, 604}

\bibitem[\protect\citeauthoryear{{Diaz}, {Bekki}, {Forbes}, {Couch},
  {Drinkwater}  \& {Deeley}}{{Diaz} et~al.}{2018}]{Diaz2018}
{Diaz} J.,  {Bekki} K.,  {Forbes} D.~A.,  {Couch} W.~J.,  {Drinkwater} M.~J.,
  {Deeley} S.,  2018, \mn@doi [\mnras] {10.1093/mnras/sty743}, \href
  {https://ui.adsabs.harvard.edu/abs/2018MNRAS.477.2030D} {477, 2030}

\bibitem[\protect\citeauthoryear{{Diemer}}{{Diemer}}{2018}]{Diemer2018ApJS}
{Diemer} B.,  2018, \mn@doi [\apjs] {10.3847/1538-4365/aaee8c}, \href
  {https://ui.adsabs.harvard.edu/abs/2018ApJS..239...35D} {239, 35}

\bibitem[\protect\citeauthoryear{{Diemer}, {Sparre}, {Abramson}  \&
  {Torrey}}{{Diemer} et~al.}{2017}]{Diemer2017}
{Diemer} B.,  {Sparre} M.,  {Abramson} L.~E.,   {Torrey} P.,  2017, \mn@doi
  [\apj] {10.3847/1538-4357/aa68e5}, \href
  {https://ui.adsabs.harvard.edu/abs/2017ApJ...839...26D} {839, 26}

\bibitem[\protect\citeauthoryear{{Diemer} et~al.,}{{Diemer}
  et~al.}{2018}]{Diemer2018ApJSHydrotools}
{Diemer} B.,  et~al., 2018, \mn@doi [\apjs] {10.3847/1538-4365/aae387}, \href
  {https://ui.adsabs.harvard.edu/abs/2018ApJS..238...33D} {238, 33}

\bibitem[\protect\citeauthoryear{{Donnari} et~al.,}{{Donnari}
  et~al.}{2019}]{Donnari2019}
{Donnari} M.,  et~al., 2019, \mn@doi [\mnras] {10.1093/mnras/stz712}, \href
  {https://ui.adsabs.harvard.edu/abs/2019MNRAS.485.4817D} {485, 4817}

\bibitem[\protect\citeauthoryear{{Donnari} et~al.,}{{Donnari}
  et~al.}{2021a}]{Donnari2021a}
{Donnari} M.,  et~al., 2021a, \mn@doi [\mnras] {10.1093/mnras/staa3006}, \href
  {https://ui.adsabs.harvard.edu/abs/2021MNRAS.500.4004D} {500, 4004}

\bibitem[\protect\citeauthoryear{{Donnari}, {Pillepich}, {Nelson}, {Marinacci},
  {Vogelsberger}  \& {Hernquist}}{{Donnari} et~al.}{2021b}]{Donnari2021b}
{Donnari} M.,  {Pillepich} A.,  {Nelson} D.,  {Marinacci} F.,  {Vogelsberger}
  M.,   {Hernquist} L.,  2021b, \mn@doi [\mnras] {10.1093/mnras/stab1950},
  \href {https://ui.adsabs.harvard.edu/abs/2021MNRAS.506.4760D} {506, 4760}

\bibitem[\protect\citeauthoryear{{Du}, {Ho}, {Debattista}, {Pillepich},
  {Nelson}, {Hernquist}  \& {Weinberger}}{{Du} et~al.}{2021}]{Du2021}
{Du} M.,  {Ho} L.~C.,  {Debattista} V.~P.,  {Pillepich} A.,  {Nelson} D.,
  {Hernquist} L.,   {Weinberger} R.,  2021, \mn@doi [\apj]
  {10.3847/1538-4357/ac0e98}, \href
  {https://ui.adsabs.harvard.edu/abs/2021ApJ...919..135D} {919, 135}

\bibitem[\protect\citeauthoryear{Dubois, Peirani, Pichon, Devriendt, Gavazzi,
  Welker  \& Volonteri}{Dubois et~al.}{2016}]{Dubois2016TheFeedback}
Dubois Y.,  Peirani S.,  Pichon C.,  Devriendt J.,  Gavazzi R.,  Welker C.,
  Volonteri M.,  2016, \mn@doi [\mnras] {10.1093/mnras/stw2265}, 463, 3948

\bibitem[\protect\citeauthoryear{{Ellison}, {S{\'a}nchez}, {Ibarra-Medel},
  {Antonio}, {Mendel}  \& {Barrera-Ballesteros}}{{Ellison}
  et~al.}{2018}]{Ellison2018}
{Ellison} S.~L.,  {S{\'a}nchez} S.~F.,  {Ibarra-Medel} H.,  {Antonio} B.,
  {Mendel} J.~T.,   {Barrera-Ballesteros} J.,  2018, \mn@doi [\mnras]
  {10.1093/mnras/stx2882}, \href
  {https://ui.adsabs.harvard.edu/abs/2018MNRAS.474.2039E} {474, 2039}

\bibitem[\protect\citeauthoryear{{French}, {Yang}, {Zabludoff}, {Narayanan},
  {Shirley}, {Walter}, {Smith}  \& {Tremonti}}{{French}
  et~al.}{2015}]{French2015}
{French} K.~D.,  {Yang} Y.,  {Zabludoff} A.,  {Narayanan} D.,  {Shirley} Y.,
  {Walter} F.,  {Smith} J.-D.,   {Tremonti} C.~A.,  2015, \mn@doi [\apj]
  {10.1088/0004-637X/801/1/1}, \href
  {https://ui.adsabs.harvard.edu/abs/2015ApJ...801....1F} {801, 1}

\bibitem[\protect\citeauthoryear{{Genel} et~al.,}{{Genel}
  et~al.}{2018}]{Genel2018}
{Genel} S.,  et~al., 2018, \mn@doi [\mnras] {10.1093/mnras/stx3078}, \href
  {https://ui.adsabs.harvard.edu/abs/2018MNRAS.474.3976G} {474, 3976}

\bibitem[\protect\citeauthoryear{Grand, Springel, G{\'{o}}mez, Marinacci,
  Pakmor, Campbell  \& Jenkins}{Grand et~al.}{2016}]{Grand2016VerticalContext}
Grand R.~J.,  Springel V.,  G{\'{o}}mez F.~A.,  Marinacci F.,  Pakmor R.,
  Campbell D.~J.,   Jenkins A.,  2016, \mn@doi [\mnras] {10.1093/mnras/stw601},
  459, 199

\bibitem[\protect\citeauthoryear{{Gunn} \& {Gott}}{{Gunn} \&
  {Gott}}{1972}]{Gunn1972}
{Gunn} J.~E.,  {Gott} J.~Richard I.,  1972, \mn@doi [\apj] {10.1086/151605},
  \href {https://ui.adsabs.harvard.edu/abs/1972ApJ...176....1G} {176, 1}

\bibitem[\protect\citeauthoryear{{Guo} et~al.,}{{Guo} et~al.}{2020}]{Guo2020}
{Guo} K.,  et~al., 2020, \mn@doi [\mnras] {10.1093/mnras/stz3042}, \href
  {https://ui.adsabs.harvard.edu/abs/2020MNRAS.491..773G} {491, 773}

\bibitem[\protect\citeauthoryear{{Hao}, {Shi}, {Chen}, {Xia}, {Gu}, {Guo}, {Yu}
   \& {Li}}{{Hao} et~al.}{2019}]{Hao2019}
{Hao} C.-N.,  {Shi} Y.,  {Chen} Y.,  {Xia} X.,  {Gu} Q.,  {Guo} R.,  {Yu} X.,
  {Li} S.,  2019, \mn@doi [\apjl] {10.3847/2041-8213/ab42e5}, \href
  {https://ui.adsabs.harvard.edu/abs/2019ApJ...883L..36H} {883, L36}

\bibitem[\protect\citeauthoryear{Harris et~al.,}{Harris
  et~al.}{2020}]{Harris2020NumPy-Array}
Harris C.~R.,  et~al., 2020, \mn@doi [Nature] {10.1038/s41586-020-2649-2}, 585,
  357–362

\bibitem[\protect\citeauthoryear{{Hernquist}}{{Hernquist}}{1992}]{Hernquist1992ApJ400}
{Hernquist} L.,  1992, \mn@doi [\apj] {10.1086/172009}, \href
  {https://ui.adsabs.harvard.edu/abs/1992ApJ...400..460H} {400, 460}

\bibitem[\protect\citeauthoryear{{Hernquist}}{{Hernquist}}{1993}]{Hernquist1993ApJ409}
{Hernquist} L.,  1993, \mn@doi [\apj] {10.1086/172686}, \href
  {https://ui.adsabs.harvard.edu/abs/1993ApJ...409..548H} {409, 548}

\bibitem[\protect\citeauthoryear{{Hopkins}, {Hernquist}, {Cox}, {Di Matteo},
  {Robertson}  \& {Springel}}{{Hopkins} et~al.}{2006}]{Hopkins2006ApJS163}
{Hopkins} P.~F.,  {Hernquist} L.,  {Cox} T.~J.,  {Di Matteo} T.,  {Robertson}
  B.,   {Springel} V.,  2006, \mn@doi [\apjs] {10.1086/499298}, \href
  {https://ui.adsabs.harvard.edu/abs/2006ApJS..163....1H} {163, 1}

\bibitem[\protect\citeauthoryear{{Hopkins}, {Hernquist}, {Cox}  \&
  {Kere{\v{s}}}}{{Hopkins} et~al.}{2008}]{Hopkins2008ApJS175}
{Hopkins} P.~F.,  {Hernquist} L.,  {Cox} T.~J.,   {Kere{\v{s}}} D.,  2008,
  \mn@doi [\apjs] {10.1086/524362}, \href
  {https://ui.adsabs.harvard.edu/abs/2008ApJS..175..356H} {175, 356}

\bibitem[\protect\citeauthoryear{Hopkins et~al.,}{Hopkins
  et~al.}{2010}]{Hopkins2010MergersMatter}
Hopkins P.~F.,  et~al., 2010, \mn@doi [\apj] {10.1088/0004-637X/715/1/202},
  715, 202

\bibitem[\protect\citeauthoryear{{Hunt} et~al.,}{{Hunt}
  et~al.}{2018}]{Hunt2018StellarZ0.7}
{Hunt} Q.,  et~al., 2018, \mn@doi [\apjl] {10.3847/2041-8213/aaca9a}, \href
  {https://ui.adsabs.harvard.edu/abs/2018ApJ...860L..18H} {860, L18}

\bibitem[\protect\citeauthoryear{{Hunter}}{{Hunter}}{2007}]{Hunter2007}
{Hunter} J.~D.,  2007, \mn@doi [Computing in Science and Engineering]
  {10.1109/MCSE.2007.55}, \href
  {https://ui.adsabs.harvard.edu/abs/2007CSE.....9...90H} {9, 90}

\bibitem[\protect\citeauthoryear{{Joshi}, {Pillepich}, {Nelson}, {Marinacci},
  {Springel}, {Rodriguez-Gomez}, {Vogelsberger}  \& {Hernquist}}{{Joshi}
  et~al.}{2020}]{Joshi2020TheClusters}
{Joshi} G.~D.,  {Pillepich} A.,  {Nelson} D.,  {Marinacci} F.,  {Springel} V.,
  {Rodriguez-Gomez} V.,  {Vogelsberger} M.,   {Hernquist} L.,  2020, \mn@doi
  [\mnras] {10.1093/mnras/staa1668}, \href
  {https://ui.adsabs.harvard.edu/abs/2020MNRAS.496.2673J} {496, 2673}

\bibitem[\protect\citeauthoryear{{Jung}, {Choi}, {Wong}, {Kimm}, {Chung}  \&
  {Yi}}{{Jung} et~al.}{2018}]{Jung2018OnSimulations}
{Jung} S.~L.,  {Choi} H.,  {Wong} O.~I.,  {Kimm} T.,  {Chung} A.,   {Yi} S.~K.,
   2018, \mn@doi [\apj] {10.3847/1538-4357/aadda2}, \href
  {https://ui.adsabs.harvard.edu/abs/2018ApJ...865..156J} {865, 156}

\bibitem[\protect\citeauthoryear{{Kere{\v{s}}}, {Katz}, {Weinberg}  \&
  {Dav{\'e}}}{{Kere{\v{s}}} et~al.}{2005}]{Keres2005}
{Kere{\v{s}}} D.,  {Katz} N.,  {Weinberg} D.~H.,   {Dav{\'e}} R.,  2005,
  \mn@doi [\mnras] {10.1111/j.1365-2966.2005.09451.x}, \href
  {https://ui.adsabs.harvard.edu/abs/2005MNRAS.363....2K} {363, 2}

\bibitem[\protect\citeauthoryear{{Khim}, {Yi}, {Pichon}, {Dubois}, {Devriendt},
  {Choi}, {Bryant}  \& {Croom}}{{Khim} et~al.}{2021}]{Khim2021}
{Khim} D.~J.,  {Yi} S.~K.,  {Pichon} C.,  {Dubois} Y.,  {Devriendt} J.,  {Choi}
  H.,  {Bryant} J.~J.,   {Croom} S.~M.,  2021, \mn@doi [\apjs]
  {10.3847/1538-4365/abf043}, \href
  {https://ui.adsabs.harvard.edu/abs/2021ApJS..254...27K} {254, 27}

\bibitem[\protect\citeauthoryear{{Khoperskov}, {Haywood}, {Di Matteo},
  {Lehnert}  \& {Combes}}{{Khoperskov} et~al.}{2018}]{Khoperskov2018}
{Khoperskov} S.,  {Haywood} M.,  {Di Matteo} P.,  {Lehnert} M.~D.,   {Combes}
  F.,  2018, \mn@doi [\aap] {10.1051/0004-6361/201731211}, \href
  {https://ui.adsabs.harvard.edu/abs/2018A&A...609A..60K} {609, A60}

\bibitem[\protect\citeauthoryear{Lang et~al.,}{Lang
  et~al.}{2014}]{Lang2014BulgeCandels/3D-HST}
Lang P.,  et~al., 2014, \apj, 788, 11

\bibitem[\protect\citeauthoryear{{Ludlow}, {Fall}, {Schaye}  \&
  {Obreschkow}}{{Ludlow} et~al.}{2021}]{Ludlow2021}
{Ludlow} A.~D.,  {Fall} S.~M.,  {Schaye} J.,   {Obreschkow} D.,  2021, arXiv
  e-prints, \href {https://ui.adsabs.harvard.edu/abs/2021arXiv210503561L} {p.
  arXiv:2105.03561}

\bibitem[\protect\citeauthoryear{{Luo}, {Li}, {Kang}, {Li}  \& {Wang}}{{Luo}
  et~al.}{2020}]{Luo2020WhatGalaxies}
{Luo} Y.,  {Li} Z.,  {Kang} X.,  {Li} Z.,   {Wang} P.,  2020, \mn@doi [\mnras]
  {10.1093/mnrasl/slaa099}, \href
  {https://ui.adsabs.harvard.edu/abs/2020MNRAS.496L.116L} {496, L116}

\bibitem[\protect\citeauthoryear{{Marinacci} et~al.,}{{Marinacci}
  et~al.}{2018}]{Marinacci2018MNRAS480}
{Marinacci} F.,  et~al., 2018, \mn@doi [\mnras] {10.1093/mnras/sty2206}, \href
  {https://ui.adsabs.harvard.edu/abs/2018MNRAS.480.5113M} {480, 5113}

\bibitem[\protect\citeauthoryear{{Martig}, {Bournaud}, {Teyssier}  \&
  {Dekel}}{{Martig} et~al.}{2009}]{Martig2009}
{Martig} M.,  {Bournaud} F.,  {Teyssier} R.,   {Dekel} A.,  2009, \mn@doi
  [\apj] {10.1088/0004-637X/707/1/250}, \href
  {https://ui.adsabs.harvard.edu/abs/2009ApJ...707..250M} {707, 250}

\bibitem[\protect\citeauthoryear{{Masters} et~al.,}{{Masters}
  et~al.}{2010}]{Masters2010}
{Masters} K.~L.,  et~al., 2010, \mn@doi [\mnras]
  {10.1111/j.1365-2966.2010.16503.x}, \href
  {https://ui.adsabs.harvard.edu/abs/2010MNRAS.405..783M} {405, 783}

\bibitem[\protect\citeauthoryear{Moffett et~al.,}{Moffett
  et~al.}{2016}]{Moffett2016GalaxyDiscs}
Moffett A.~J.,  et~al., 2016, \mn@doi [\mnras] {10.1093/mnras/stw1861}, 462,
  4336

\bibitem[\protect\citeauthoryear{{Morselli}, {Popesso}, {Cibinel}, {Oesch},
  {Montes}, {Atek}, {Illingworth}  \& {Holden}}{{Morselli}
  et~al.}{2019}]{Morselli2019}
{Morselli} L.,  {Popesso} P.,  {Cibinel} A.,  {Oesch} P.~A.,  {Montes} M.,
  {Atek} H.,  {Illingworth} G.~D.,   {Holden} B.,  2019, \mn@doi [\aap]
  {10.1051/0004-6361/201834559}, \href
  {https://ui.adsabs.harvard.edu/abs/2019A&A...626A..61M} {626, A61}

\bibitem[\protect\citeauthoryear{{Mosleh}, {Tacchella}, {Renzini}, {Carollo},
  {Molaeinezhad}, {Onodera}, {Khosroshahi}  \& {Lilly}}{{Mosleh}
  et~al.}{2017}]{Mosleh2017}
{Mosleh} M.,  {Tacchella} S.,  {Renzini} A.,  {Carollo} C.~M.,  {Molaeinezhad}
  A.,  {Onodera} M.,  {Khosroshahi} H.~G.,   {Lilly} S.,  2017, \mn@doi [\apj]
  {10.3847/1538-4357/aa5f14}, \href
  {https://ui.adsabs.harvard.edu/abs/2017ApJ...837....2M} {837, 2}

\bibitem[\protect\citeauthoryear{{Moustakas} et~al.,}{{Moustakas}
  et~al.}{2013}]{Moustakas2013}
{Moustakas} J.,  et~al., 2013, \mn@doi [\apj] {10.1088/0004-637X/767/1/50},
  \href {https://ui.adsabs.harvard.edu/abs/2013ApJ...767...50M} {767, 50}

\bibitem[\protect\citeauthoryear{{Muldrew}, {Hatch}  \& {Cooke}}{{Muldrew}
  et~al.}{2018}]{Muldrew2018}
{Muldrew} S.~I.,  {Hatch} N.~A.,   {Cooke} E.~A.,  2018, \mn@doi [\mnras]
  {10.1093/mnras/stx2454}, \href
  {https://ui.adsabs.harvard.edu/abs/2018MNRAS.473.2335M} {473, 2335}

\bibitem[\protect\citeauthoryear{{Muzzin} et~al.,}{{Muzzin}
  et~al.}{2013}]{Muzzin2013}
{Muzzin} A.,  et~al., 2013, \mn@doi [\apj] {10.1088/0004-637X/777/1/18}, \href
  {https://ui.adsabs.harvard.edu/abs/2013ApJ...777...18M} {777, 18}

\bibitem[\protect\citeauthoryear{{Naiman} et~al.,}{{Naiman}
  et~al.}{2018}]{Naiman2018MNRAS477}
{Naiman} J.~P.,  et~al., 2018, \mn@doi [\mnras] {10.1093/mnras/sty618}, \href
  {https://ui.adsabs.harvard.edu/abs/2018MNRAS.477.1206N} {477, 1206}

\bibitem[\protect\citeauthoryear{{Nelson} et~al.,}{{Nelson}
  et~al.}{2016}]{Nelson2016}
{Nelson} E.~J.,  et~al., 2016, \mn@doi [\apj] {10.3847/0004-637X/828/1/27},
  \href {https://ui.adsabs.harvard.edu/abs/2016ApJ...828...27N} {828, 27}

\bibitem[\protect\citeauthoryear{Nelson et~al.,}{Nelson
  et~al.}{2018}]{Nelson2018FirstBimodality}
Nelson D.,  et~al., 2018, \mn@doi [\mnras] {10.1093/mnras/stx3040}, 475, 624

\bibitem[\protect\citeauthoryear{{Nelson} et~al.,}{{Nelson}
  et~al.}{2019a}]{Nelson2019ComAC}
{Nelson} D.,  et~al., 2019a, \mn@doi [Computational Astrophysics and Cosmology]
  {10.1186/s40668-019-0028-x}, \href
  {https://ui.adsabs.harvard.edu/abs/2019ComAC...6....2N} {6, 2}

\bibitem[\protect\citeauthoryear{{Nelson} et~al.,}{{Nelson}
  et~al.}{2019b}]{Nelson2019}
{Nelson} D.,  et~al., 2019b, \mn@doi [\mnras] {10.1093/mnras/stz2306}, \href
  {https://ui.adsabs.harvard.edu/abs/2019MNRAS.490.3234N} {490, 3234}

\bibitem[\protect\citeauthoryear{{Nelson} et~al.,}{{Nelson}
  et~al.}{2021}]{Nelson2021}
{Nelson} E.~J.,  et~al., 2021, \mn@doi [\mnras] {10.1093/mnras/stab2131}, \href
  {https://ui.adsabs.harvard.edu/abs/2021MNRAS.tmp.2068N} {}

\bibitem[\protect\citeauthoryear{Noguchi}{Noguchi}{1999}]{Noguchi1999EarlyDisks}
Noguchi M.,  1999, \mn@doi [\apj] {10.1086/306932}, 514, 77

\bibitem[\protect\citeauthoryear{{Osborne} et~al.,}{{Osborne}
  et~al.}{2020}]{Osborne2020}
{Osborne} C.,  et~al., 2020, \mn@doi [\apj] {10.3847/1538-4357/abb5af}, \href
  {https://ui.adsabs.harvard.edu/abs/2020ApJ...902...77O} {902, 77}

\bibitem[\protect\citeauthoryear{{Pandya} et~al.,}{{Pandya}
  et~al.}{2017}]{Pandya2017}
{Pandya} V.,  et~al., 2017, \mn@doi [\mnras] {10.1093/mnras/stx2027}, \href
  {https://ui.adsabs.harvard.edu/abs/2017MNRAS.472.2054P} {472, 2054}

\bibitem[\protect\citeauthoryear{{Park} et~al.,}{{Park}
  et~al.}{2019}]{Park2019}
{Park} M.-J.,  et~al., 2019, \mn@doi [\apj] {10.3847/1538-4357/ab3afe}, \href
  {https://ui.adsabs.harvard.edu/abs/2019ApJ...883...25P} {883, 25}

\bibitem[\protect\citeauthoryear{{Pathak}, {Belli}  \& {Weinberger}}{{Pathak}
  et~al.}{2021}]{Pathak2021}
{Pathak} D.,  {Belli} S.,   {Weinberger} R.,  2021, \mn@doi [\apjl]
  {10.3847/2041-8213/ac13a7}, \href
  {https://ui.adsabs.harvard.edu/abs/2021ApJ...916L..23P} {916, L23}

\bibitem[\protect\citeauthoryear{{Peng} et~al.,}{{Peng}
  et~al.}{2010}]{Peng2010MassFunction}
{Peng} Y.-j.,  et~al., 2010, \mn@doi [\apj] {10.1088/0004-637X/721/1/193},
  \href {https://ui.adsabs.harvard.edu/abs/2010ApJ...721..193P} {721, 193}

\bibitem[\protect\citeauthoryear{{Pillepich} et~al.,}{{Pillepich}
  et~al.}{2018a}]{Pillepich2018TNGmodels}
{Pillepich} A.,  et~al., 2018a, \mn@doi [\mnras] {10.1093/mnras/stx2656}, \href
  {https://ui.adsabs.harvard.edu/abs/2018MNRAS.473.4077P} {473, 4077}

\bibitem[\protect\citeauthoryear{Pillepich et~al.,}{Pillepich
  et~al.}{2018b}]{Pillepich2018FirstGalaxies}
Pillepich A.,  et~al., 2018b, \mn@doi [\mnras] {10.1093/mnras/stx3112}, 475,
  648

\bibitem[\protect\citeauthoryear{{Pillepich} et~al.,}{{Pillepich}
  et~al.}{2019}]{Pillepich2019}
{Pillepich} A.,  et~al., 2019, \mn@doi [\mnras] {10.1093/mnras/stz2338}, \href
  {https://ui.adsabs.harvard.edu/abs/2019MNRAS.490.3196P} {490, 3196}

\bibitem[\protect\citeauthoryear{{Pillepich}, {Nelson}, {Truong}, {Weinberger},
  {Martin-Navarro}, {Springel}, {Faber}  \& {Hernquist}}{{Pillepich}
  et~al.}{2021}]{Pillepich2021}
{Pillepich} A.,  {Nelson} D.,  {Truong} N.,  {Weinberger} R.,  {Martin-Navarro}
  I.,  {Springel} V.,  {Faber} S.~M.,   {Hernquist} L.,  2021, \mn@doi [\mnras]
  {10.1093/mnras/stab2779}, \href
  {https://ui.adsabs.harvard.edu/abs/2021MNRAS.508.4667P} {508, 4667}

\bibitem[\protect\citeauthoryear{{Quai}, {Hani}, {Ellison}, {Patton}  \&
  {Woo}}{{Quai} et~al.}{2021}]{Quai2021}
{Quai} S.,  {Hani} M.~H.,  {Ellison} S.~L.,  {Patton} D.~R.,   {Woo} J.,  2021,
  \mn@doi [\mnras] {10.1093/mnras/stab988}, \href
  {https://ui.adsabs.harvard.edu/abs/2021MNRAS.504.1888Q} {504, 1888}

\bibitem[\protect\citeauthoryear{{Renzini} \& {Peng}}{{Renzini} \&
  {Peng}}{2015}]{Renzini2015}
{Renzini} A.,  {Peng} Y.-j.,  2015, \mn@doi [\apjl]
  {10.1088/2041-8205/801/2/L29}, \href
  {https://ui.adsabs.harvard.edu/abs/2015ApJ...801L..29R} {801, L29}

\bibitem[\protect\citeauthoryear{{Rhee}, {Smith}, {Choi}, {Contini}, {Jung},
  {Han}  \& {Yi}}{{Rhee} et~al.}{2020}]{Rhee2020}
{Rhee} J.,  {Smith} R.,  {Choi} H.,  {Contini} E.,  {Jung} S.~L.,  {Han} S.,
  {Yi} S.~K.,  2020, \mn@doi [\apjs] {10.3847/1538-4365/ab7377}, \href
  {https://ui.adsabs.harvard.edu/abs/2020ApJS..247...45R} {247, 45}

\bibitem[\protect\citeauthoryear{Rodriguez-Gomez et~al.,}{Rodriguez-Gomez
  et~al.}{2016}]{Rodriguez-Gomez2016TheStars}
Rodriguez-Gomez V.,  et~al., 2016, \mn@doi [\mnras] {10.1093/mnras/stw456},
  458, 2371

\bibitem[\protect\citeauthoryear{{Rodriguez-Gomez} et~al.,}{{Rodriguez-Gomez}
  et~al.}{2019}]{Rodriguez-Gomez2019}
{Rodriguez-Gomez} V.,  et~al., 2019, \mn@doi [\mnras] {10.1093/mnras/sty3345},
  \href {https://ui.adsabs.harvard.edu/abs/2019MNRAS.483.4140R} {483, 4140}

\bibitem[\protect\citeauthoryear{{Rodr{\'\i}guez Montero}, {Dav{\'e}}, {Wild},
  {Angl{\'e}s-Alc{\'a}zar}  \& {Narayanan}}{{Rodr{\'\i}guez Montero}
  et~al.}{2019}]{Rodriguez-Montero2019}
{Rodr{\'\i}guez Montero} F.,  {Dav{\'e}} R.,  {Wild} V.,
  {Angl{\'e}s-Alc{\'a}zar} D.,   {Narayanan} D.,  2019, \mn@doi [\mnras]
  {10.1093/mnras/stz2580}, \href
  {https://ui.adsabs.harvard.edu/abs/2019MNRAS.490.2139R} {490, 2139}

\bibitem[\protect\citeauthoryear{{Sales}, {Navarro}, {Theuns}, {Schaye},
  {White}, {Frenk}, {Crain}  \& {Dalla Vecchia}}{{Sales}
  et~al.}{2012}]{Sales2012}
{Sales} L.~V.,  {Navarro} J.~F.,  {Theuns} T.,  {Schaye} J.,  {White} S. D.~M.,
   {Frenk} C.~S.,  {Crain} R.~A.,   {Dalla Vecchia} C.,  2012, \mn@doi [\mnras]
  {10.1111/j.1365-2966.2012.20975.x}, \href
  {https://ui.adsabs.harvard.edu/abs/2012MNRAS.423.1544S} {423, 1544}

\bibitem[\protect\citeauthoryear{{Schawinski} et~al.,}{{Schawinski}
  et~al.}{2014}]{Schawinski2014}
{Schawinski} K.,  et~al., 2014, \mn@doi [\mnras] {10.1093/mnras/stu327}, \href
  {https://ui.adsabs.harvard.edu/abs/2014MNRAS.440..889S} {440, 889}

\bibitem[\protect\citeauthoryear{{Silk} \& {Rees}}{{Silk} \&
  {Rees}}{1998}]{Silk1998}
{Silk} J.,  {Rees} M.~J.,  1998, \aap, \href
  {https://ui.adsabs.harvard.edu/abs/1998A&A...331L...1S} {331, L1}

\bibitem[\protect\citeauthoryear{{Smethurst} et~al.,}{{Smethurst}
  et~al.}{2018}]{Smethurst2018}
{Smethurst} R.~J.,  et~al., 2018, \mn@doi [\mnras] {10.1093/mnras/stx2547},
  \href {https://ui.adsabs.harvard.edu/abs/2018MNRAS.473.2679S} {473, 2679}

\bibitem[\protect\citeauthoryear{{Somerville} \& {Dav{\'e}}}{{Somerville} \&
  {Dav{\'e}}}{2015}]{Somerville2015}
{Somerville} R.~S.,  {Dav{\'e}} R.,  2015, \mn@doi [\araa]
  {10.1146/annurev-astro-082812-140951}, \href
  {https://ui.adsabs.harvard.edu/abs/2015ARA&A..53...51S} {53, 51}

\bibitem[\protect\citeauthoryear{{Somerville}, {Hopkins}, {Cox}, {Robertson}
  \& {Hernquist}}{{Somerville} et~al.}{2008}]{Somerville2008}
{Somerville} R.~S.,  {Hopkins} P.~F.,  {Cox} T.~J.,  {Robertson} B.~E.,
  {Hernquist} L.,  2008, \mn@doi [\mnras] {10.1111/j.1365-2966.2008.13805.x},
  \href {https://ui.adsabs.harvard.edu/abs/2008MNRAS.391..481S} {391, 481}

\bibitem[\protect\citeauthoryear{{Springel} \& {Hernquist}}{{Springel} \&
  {Hernquist}}{2003}]{Springel2003MNRAS339}
{Springel} V.,  {Hernquist} L.,  2003, \mn@doi [\mnras]
  {10.1046/j.1365-8711.2003.06207.x}, \href
  {https://ui.adsabs.harvard.edu/abs/2003MNRAS.339..312S} {339, 312}

\bibitem[\protect\citeauthoryear{{Springel}, {Di Matteo}  \&
  {Hernquist}}{{Springel} et~al.}{2005}]{Springel2005BH}
{Springel} V.,  {Di Matteo} T.,   {Hernquist} L.,  2005, \mn@doi [\apjl]
  {10.1086/428772}, \href
  {https://ui.adsabs.harvard.edu/abs/2005ApJ...620L..79S} {620, L79}

\bibitem[\protect\citeauthoryear{{Springel} et~al.,}{{Springel}
  et~al.}{2018}]{Springel2018MNRAS475}
{Springel} V.,  et~al., 2018, \mn@doi [\mnras] {10.1093/mnras/stx3304}, \href
  {https://ui.adsabs.harvard.edu/abs/2018MNRAS.475..676S} {475, 676}

\bibitem[\protect\citeauthoryear{{Strateva} et~al.,}{{Strateva}
  et~al.}{2001}]{Strateva2001}
{Strateva} I.,  et~al., 2001, \mn@doi [\aj] {10.1086/323301}, \href
  {https://ui.adsabs.harvard.edu/abs/2001AJ....122.1861S} {122, 1861}

\bibitem[\protect\citeauthoryear{{Suess}, {Bezanson}, {Spilker}, {Kriek},
  {Greene}, {Feldmann}, {Hunt}  \& {Narayanan}}{{Suess}
  et~al.}{2017}]{Suess2017}
{Suess} K.~A.,  {Bezanson} R.,  {Spilker} J.~S.,  {Kriek} M.,  {Greene} J.~E.,
  {Feldmann} R.,  {Hunt} Q.,   {Narayanan} D.,  2017, \mn@doi [\apjl]
  {10.3847/2041-8213/aa85dc}, \href
  {https://ui.adsabs.harvard.edu/abs/2017ApJ...846L..14S} {846, L14}

\bibitem[\protect\citeauthoryear{{Tacchella} et~al.,}{{Tacchella}
  et~al.}{2015}]{Tacchella2015Sci}
{Tacchella} S.,  et~al., 2015, \mn@doi [Science] {10.1126/science.1261094},
  \href {https://ui.adsabs.harvard.edu/abs/2015Sci...348..314T} {348, 314}

\bibitem[\protect\citeauthoryear{{Tacchella}, {Dekel}, {Carollo}, {Ceverino},
  {DeGraf}, {Lapiner}, {Mandelker}  \& {Primack Joel}}{{Tacchella}
  et~al.}{2016a}]{Tacchella2016Thereplenishment}
{Tacchella} S.,  {Dekel} A.,  {Carollo} C.~M.,  {Ceverino} D.,  {DeGraf} C.,
  {Lapiner} S.,  {Mandelker} N.,   {Primack Joel} R.,  2016a, \mn@doi [\mnras]
  {10.1093/mnras/stw131}, \href
  {https://ui.adsabs.harvard.edu/abs/2016MNRAS.457.2790T} {457, 2790}

\bibitem[\protect\citeauthoryear{{Tacchella}, {Dekel}, {Carollo}, {Ceverino},
  {DeGraf}, {Lapiner}, {Mandelker}  \& {Primack}}{{Tacchella}
  et~al.}{2016b}]{Tacchella2016}
{Tacchella} S.,  {Dekel} A.,  {Carollo} C.~M.,  {Ceverino} D.,  {DeGraf} C.,
  {Lapiner} S.,  {Mandelker} N.,   {Primack} J.~R.,  2016b, \mn@doi [\mnras]
  {10.1093/mnras/stw303}, \href
  {https://ui.adsabs.harvard.edu/abs/2016MNRAS.458..242T} {458, 242}

\bibitem[\protect\citeauthoryear{{Tacchella} et~al.,}{{Tacchella}
  et~al.}{2018}]{Tacchella2018}
{Tacchella} S.,  et~al., 2018, \mn@doi [\apj] {10.3847/1538-4357/aabf8b}, \href
  {https://ui.adsabs.harvard.edu/abs/2018ApJ...859...56T} {859, 56}

\bibitem[\protect\citeauthoryear{{Tacchella} et~al.,}{{Tacchella}
  et~al.}{2019}]{Tacchella2019}
{Tacchella} S.,  et~al., 2019, \mn@doi [\mnras] {10.1093/mnras/stz1657}, \href
  {https://ui.adsabs.harvard.edu/abs/2019MNRAS.487.5416T} {487, 5416}

\bibitem[\protect\citeauthoryear{{Tacchella} et~al.,}{{Tacchella}
  et~al.}{2021}]{Tacchella2021}
{Tacchella} S.,  et~al., 2021, arXiv e-prints, \href
  {https://ui.adsabs.harvard.edu/abs/2021arXiv210212494T} {p. arXiv:2102.12494}

\bibitem[\protect\citeauthoryear{{Terrazas} et~al.,}{{Terrazas}
  et~al.}{2020}]{Terrazas2020}
{Terrazas} B.~A.,  et~al., 2020, \mn@doi [\mnras] {10.1093/mnras/staa374},
  \href {https://ui.adsabs.harvard.edu/abs/2020MNRAS.493.1888T} {493, 1888}

\bibitem[\protect\citeauthoryear{{Toomre}}{{Toomre}}{1977}]{Toomre1977}
{Toomre} A.,  1977, in {Tinsley} B.~M.,  {Larson} Richard B.~Gehret D.~C.,
  eds, Evolution of Galaxies and Stellar Populations. p.~401

\bibitem[\protect\citeauthoryear{{Trussler}, {Maiolino}, {Maraston}, {Peng},
  {Thomas}, {Goddard}  \& {Lian}}{{Trussler} et~al.}{2020}]{Trussler2020}
{Trussler} J.,  {Maiolino} R.,  {Maraston} C.,  {Peng} Y.,  {Thomas} D.,
  {Goddard} D.,   {Lian} J.,  2020, \mn@doi [\mnras] {10.1093/mnras/stz3286},
  \href {https://ui.adsabs.harvard.edu/abs/2020MNRAS.491.5406T} {491, 5406}

\bibitem[\protect\citeauthoryear{Vogelsberger et~al.,}{Vogelsberger
  et~al.}{2014a}]{Vogelsberger2014IntroducingUniverse}
Vogelsberger M.,  et~al., 2014a, \mn@doi [\mnras] {10.1093/mnras/stu1536}, 444,
  1518

\bibitem[\protect\citeauthoryear{{Vogelsberger} et~al.,}{{Vogelsberger}
  et~al.}{2014b}]{Vogelsberger2014Natur509}
{Vogelsberger} M.,  et~al., 2014b, \mn@doi [\nat] {10.1038/nature13316}, \href
  {https://ui.adsabs.harvard.edu/abs/2014Natur.509..177V} {509, 177}

\bibitem[\protect\citeauthoryear{{Vogelsberger}, {Marinacci}, {Torrey}  \&
  {Puchwein}}{{Vogelsberger} et~al.}{2020}]{Vogelsberger2020Nat2}
{Vogelsberger} M.,  {Marinacci} F.,  {Torrey} P.,   {Puchwein} E.,  2020,
  \mn@doi [Nature Reviews Physics] {10.1038/s42254-019-0127-2}, \href
  {https://ui.adsabs.harvard.edu/abs/2020NatRP...2...42V} {2, 42}

\bibitem[\protect\citeauthoryear{{Weinberger} et~al.,}{{Weinberger}
  et~al.}{2017}]{Weinberger2017MNRAS465}
{Weinberger} R.,  et~al., 2017, \mn@doi [\mnras] {10.1093/mnras/stw2944}, \href
  {https://ui.adsabs.harvard.edu/abs/2017MNRAS.465.3291W} {465, 3291}

\bibitem[\protect\citeauthoryear{{Weinberger} et~al.,}{{Weinberger}
  et~al.}{2018}]{Weinberger2018}
{Weinberger} R.,  et~al., 2018, \mn@doi [\mnras] {10.1093/mnras/sty1733}, \href
  {https://ui.adsabs.harvard.edu/abs/2018MNRAS.479.4056W} {479, 4056}

\bibitem[\protect\citeauthoryear{{Wetzel}, {Tinker}, {Conroy}  \& {van den
  Bosch}}{{Wetzel} et~al.}{2013}]{Wetzel2013}
{Wetzel} A.~R.,  {Tinker} J.~L.,  {Conroy} C.,   {van den Bosch} F.~C.,  2013,
  \mn@doi [\mnras] {10.1093/mnras/stt469}, \href
  {https://ui.adsabs.harvard.edu/abs/2013MNRAS.432..336W} {432, 336}

\bibitem[\protect\citeauthoryear{{Whitaker}, {van Dokkum}, {Brammer}  \&
  {Franx}}{{Whitaker} et~al.}{2012}]{Whitaker2012}
{Whitaker} K.~E.,  {van Dokkum} P.~G.,  {Brammer} G.,   {Franx} M.,  2012,
  \mn@doi [\apjl] {10.1088/2041-8205/754/2/L29}, \href
  {https://ui.adsabs.harvard.edu/abs/2012ApJ...754L..29W} {754, L29}

\bibitem[\protect\citeauthoryear{{Whitaker} et~al.,}{{Whitaker}
  et~al.}{2017}]{Whitaker2017}
{Whitaker} K.~E.,  et~al., 2017, \mn@doi [\apj] {10.3847/1538-4357/aa6258},
  \href {https://ui.adsabs.harvard.edu/abs/2017ApJ...838...19W} {838, 19}

\bibitem[\protect\citeauthoryear{{White} \& {Rees}}{{White} \&
  {Rees}}{1978}]{White1978}
{White} S.~D.~M.,  {Rees} M.~J.,  1978, \mn@doi [\mnras]
  {10.1093/mnras/183.3.341}, \href
  {https://ui.adsabs.harvard.edu/abs/1978MNRAS.183..341W} {183, 341}

\bibitem[\protect\citeauthoryear{{Wild} et~al.,}{{Wild}
  et~al.}{2020}]{Wild2020}
{Wild} V.,  et~al., 2020, \mn@doi [\mnras] {10.1093/mnras/staa674}, \href
  {https://ui.adsabs.harvard.edu/abs/2020MNRAS.494..529W} {494, 529}

\bibitem[\protect\citeauthoryear{{Wu} et~al.,}{{Wu} et~al.}{2018}]{Wu2018}
{Wu} P.-F.,  et~al., 2018, \mn@doi [\apj] {10.3847/1538-4357/aae822}, \href
  {https://ui.adsabs.harvard.edu/abs/2018ApJ...868...37W} {868, 37}

\bibitem[\protect\citeauthoryear{{Wuyts} et~al.,}{{Wuyts}
  et~al.}{2011}]{Wuyts2011}
{Wuyts} S.,  et~al., 2011, \mn@doi [\apj] {10.1088/0004-637X/742/2/96}, \href
  {https://ui.adsabs.harvard.edu/abs/2011ApJ...742...96W} {742, 96}

\bibitem[\protect\citeauthoryear{{Yun} et~al.,}{{Yun}
  et~al.}{2019}]{Yun2019MNRAS483}
{Yun} K.,  et~al., 2019, \mn@doi [\mnras] {10.1093/mnras/sty3156}, \href
  {https://ui.adsabs.harvard.edu/abs/2019MNRAS.483.1042Y} {483, 1042}

\bibitem[\protect\citeauthoryear{{Zinger} et~al.,}{{Zinger}
  et~al.}{2020}]{Zinger2020}
{Zinger} E.,  et~al., 2020, \mn@doi [\mnras] {10.1093/mnras/staa2607}, \href
  {https://ui.adsabs.harvard.edu/abs/2020MNRAS.499..768Z} {499, 768}

\bibitem[\protect\citeauthoryear{Zolotov et~al.,}{Zolotov
  et~al.}{2015}]{Zolotov2015CompactionNuggets}
Zolotov A.,  et~al., 2015, \mn@doi [\mnras] {10.1093/mnras/stv740}, 450, 2327

\bibitem[\protect\citeauthoryear{{van Dokkum} et~al.,}{{van Dokkum}
  et~al.}{2014}]{vanDokkum2014}
{van Dokkum} P.~G.,  et~al., 2014, \mn@doi [\apj] {10.1088/0004-637X/791/1/45},
  \href {https://ui.adsabs.harvard.edu/abs/2014ApJ...791...45V} {791, 45}

\makeatother
\end{thebibliography}

%%%%%%%%%%%%%%%%%%%%%%%%%%%%%%%%%%%%%%%%%%%%%%%%%%

% Don't change these lines
\bsp	% typesetting comment
\label{lastpage}
\end{document}